\documentclass[sigconf]{acmart}
\AtBeginDocument{%
  \providecommand\BibTeX{{%
    \normalfont B\kern-0.5em{\scshape i\kern-0.25em b}\kern-0.8em\TeX}}}

\copyrightyear{2024} 
\acmYear{2024} 
\setcopyright{acmlicensed}\acmConference[CHI '24]{Proceedings of the CHI Conference on Human Factors in Computing Systems}{May 11--16, 2024}{Honolulu, HI, USA}
\acmBooktitle{Proceedings of the CHI Conference on Human Factors in Computing Systems (CHI '24), May 11--16, 2024, Honolulu, HI, USA}
\acmDOI{10.1145/3613904.3642473}
\acmISBN{979-8-4007-0330-0/24/05}

\usepackage{color, colortbl}
\usepackage{subcaption, multirow}
\usepackage{xcolor}
\usepackage{amsmath}
\usepackage{acmart-taps}
\definecolor{light-gray}{gray}{0.95}
\definecolor{airforceblue}{rgb}{0.36, 0.54, 0.66}
\definecolor{majorelleblue}{rgb}{0.38, 0.31, 0.86}
\definecolor{oceanboatblue}{rgb}{0.0, 0.47, 0.75}
\newcommand{\code}[1]{\textcolor{oceanboatblue}{\texttt{#1}}}

\begin{document}

\title{A Browser Extension for in-place Signaling and Assessment of Misinformation}


\author{Farnaz Jahanbakhsh}
\authornote{Research conducted while at MIT.}
\affiliation{%
  \institution{Stanford University}
  \city{Stanford}
  \country{USA}
}

\author{David R. Karger}
\affiliation{%
  \institution{Computer Science and Artificial Intelligence Laboratory, Massachusetts Institute of Technology}
  \city{Cambridge}
  \country{USA}
}

\renewcommand{\shortauthors}{Jahanbakhsh and Karger}

\begin{abstract}

The status-quo of misinformation moderation is a central authority, usually social platforms, deciding what content constitutes misinformation and how it should be handled. However, to preserve users’ autonomy, researchers have explored democratized misinformation moderation. One proposition is to enable users to assess content accuracy and specify whose assessments they trust. We explore how these affordances can be provided on the web, without cooperation from the platforms where users consume content.
We present a browser extension that empowers users to assess the accuracy of any content on the web and shows the user assessments from their trusted sources in-situ. Through a two-week user study, we report on how users perceive such a tool, the kind of content users want to assess, and the rationales they use in their assessments. We identify implications for designing tools that enable users to moderate content for themselves with the help of those they trust.

\end{abstract}



\begin{teaserfigure}
\begin{minipage}{.48\textwidth}
  \includegraphics[width=\textwidth]{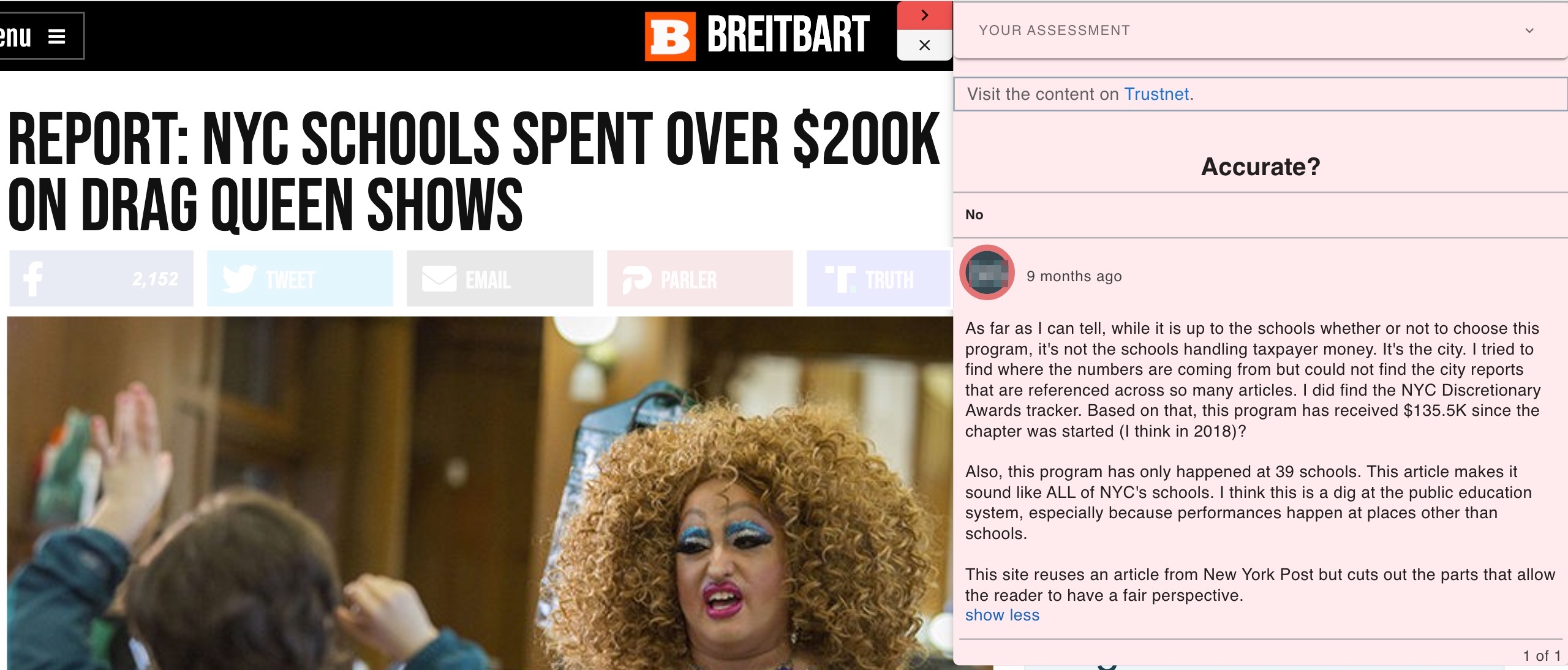}
\end{minipage}
\qquad
\begin{minipage}{.48\textwidth}
  \includegraphics[width=\textwidth]{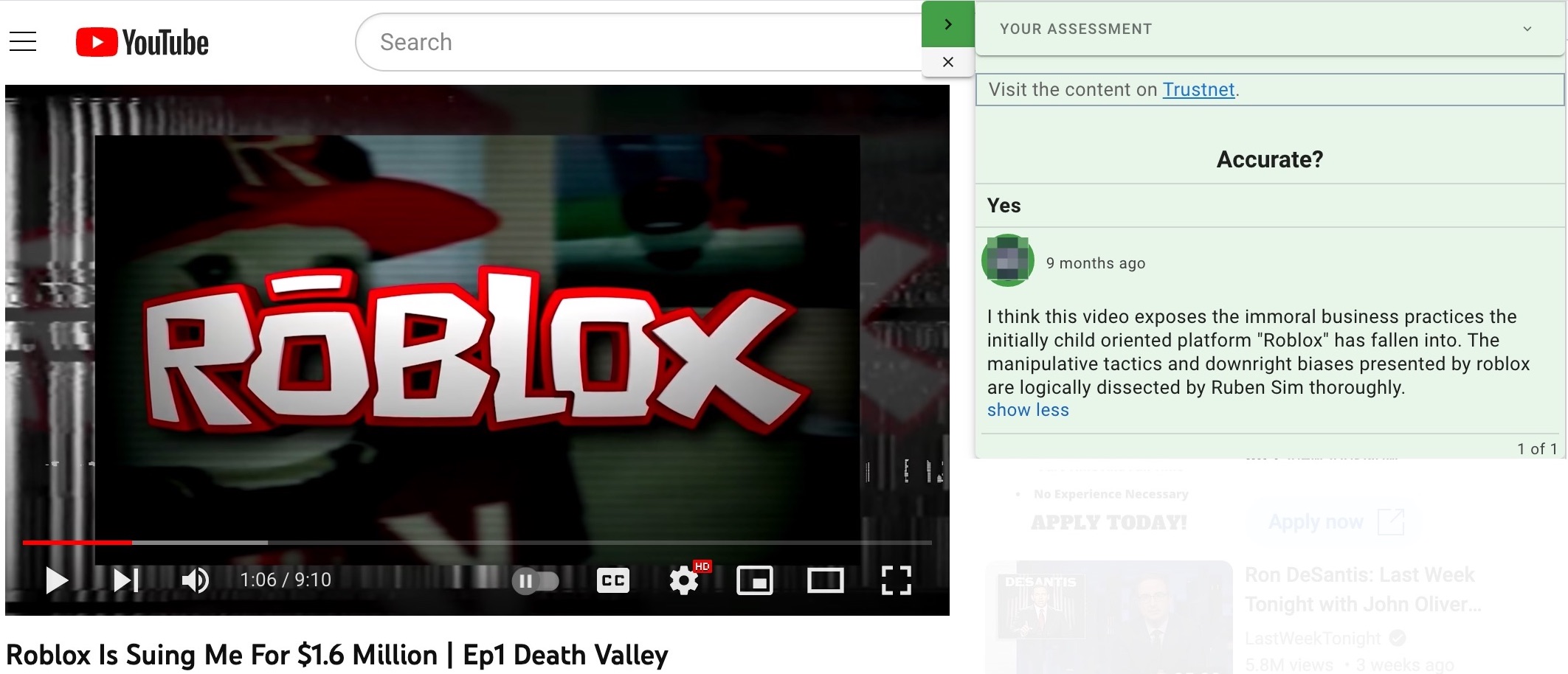}
\end{minipage}
  \caption{Upon visiting a webpage, the Trustnet extension opens a pane on the side showing the assessments of the page by those the user follows or trusts. The figure shows a news article (left) and a Youtube video (right) assessed by users of the study.}
  \Description{The image on the left shows an article from Breitbart with the headline ``Report: NYC Schools Spent over \$200K on Drag Queen Shows''. The assessments pane from our browser extension is open on the righthand side of the page. It has a light red background color indicating that the content has been assessed as inaccurate by the assessors of the user viewing the page. Inside the pane, there is a segment for the user to submit their own assessment, followed by a section where assessments of others are displayed, along with who the source of each assessment is as well as a timestamp. The image on the right shows a Youtube vide titled ``Roblox is Suing Me for \$16 Million | Ep1 Death Vallely''. The assessments pane from the Trustnet browser extension is open on the righthand side of the page. It has a light green background color indicating that the content has been assessed as accurate by the assessors of the user viewing the page. A user's assessment is shown in the pane.}
  \label{fig:teaser}
\end{teaserfigure}

\maketitle

\section{Introduction}


%

The convenience of spreading and consuming misinformation on social media platforms~\cite{wang2019systematic, vosoughi2018spread} and the media spotlight given to the havoc that misinformation has wrought in recent years~\cite{argentino2021qanon, abortionMisinfo, bengali2019whatsapp, dixit2018whatsapp} have caused a surge of attention to this age-old problem.
Researchers as well as platform operators are increasingly investing effort in combating misinformation. 
At present, the dominant approach to fighting misinformation is to place the onus for doing so on the large social platforms where many users spend their time. This approach places a critical social decision in the hands of for-profit companies, and limits users' autonomy in deciding who they trust to declare what is accurate~\cite{jahanbakhsh2022leveraging}. It also does nothing to combat misinformation that users encounter off the social platforms, including the many news sites found on the web.
The downsides of centralized misinformation moderation have motivated some researchers and practitioners to pursue decentralized solutions to the problem. For instance, Twitter Community Notes is an example of such an approach which involves all the users of the community in assessing the accuracy of content and rating each other's assessments. However, the final decision of what notes to eventually show to the public is taken in a centralized fashion, by Twitter's ``bridging algorithm"~\cite{diversityTwitterCommunityNotes}.

In contrast, a body of work has advocated for giving users more autonomy while aiding them by providing the tools that they need for discerning truth from falsehoods. For instance, it is reported that enabling users to assess the accuracy of content can reduce their likelihood to share misinformation~\cite{jahanbakhsh2021exploring}, as it nudges users to have accuracy on top of their mind~\cite{pennycook2020fighting, pennycook2021shifting}. If user assessments are captured as structured metadata, they can be displayed on content in structured form as well, similar to but separate from likes or comments, and have the potential to warn the assessor's social circle should they encounter the misinforming content~\cite{jahanbakhsh2022leveraging}. Jahanbakhsh et al. argue that users should additionally be given the ability to specify whose assessments they would like to see---in essence, leaving it to each user to decide who they trust as their moderators. They report that even on the current platforms, users do ask those they trust for their assessments; however, they do this without platform support, and in fact despite the platforms getting in their way. In a user study of a prototype social news sharing platform that offers these affordances, the authors determine that users want them and are capable of using them~\cite{jahanbakhsh2022leveraging}. 
Capturing assessments and trust relationships in structured form can be a first step in designing interventions that point users to fact-checking information from those that they trust.

To increase user autonomy, platform operators can be urged to incorporate the affordances of user generated assessments and specified trusted sources. However, this incorporation is unlikely to happen without activism or legislation, because changing the user experience drastically, particularly ceding too much control over content to their users, can impact platform revenues. 
Researchers could build a new social platform where they offer these affordances to users, for instance similar to the prototype platform presented by Jahanbakhsh et al.~\cite{jahanbakhsh2022leveraging}. But such undertaking realistically may not succeed because first, it costs much development and maintenance for this new platform to have a chance to compete with the existing social platforms; and second, users are unlikely to abandon their social network on the existing platforms and join a new one without a critical mass of their connections already present there. 

While social media platforms have been an important outlet for the circulation of misinformation, they are far from being the only one~\cite{fourney2017geographic}. 
Efforts concentrated on getting \emph{social platforms} to moderate content ignore that users consume content, by choice or by accident, outside these platforms as well. Users for instance, may encounter links to unverified or misinforming content on a news or media website that crosslinks to another, in an email newsletter, or on a news aggregator. It is unrealistic to expect that every website or platform on the web, some the very perpetrators of misinformation, offer content moderation and in a fair and rigorous way.

The pursuit for ubiquitous content moderation that preserves user autonomy is the motivation behind this work.
What follows is a case study of how enabling users to moderate content, specifically in the context of misinformation, can be deployed \emph{everywhere on the web}, in a platform-agnostic manner and without support from the social platforms or individual web sites. We use the term ``moderation'' as a concept that involves categorizing the status of content by applying labels, including those that are commonly used by platforms such as ``misleading" or ``explicit'' as well as any other label that can help a user customize their web experience, such as ``depressing''. The other aspect of moderation is taking an action, if any, on content that has been positively or negatively labeled, e.g., whether to simply flag or remove it. These two aspects are often bundled together in the status-quo practices. In this work, we do not focus on the latter aspect and simply signal the status of content. If the categorization of content is customized according to users' needs and input, then a set of tools can exist that take the labels and act in different ways on them, or within the same tool, the choice of what to do with the content can also be deferred to the user~\cite{jahanbakhsh2022leveraging}.

 In this work, we design, deploy, and evaluate a browser extension that lets \emph{anyone} assess any content---news articles, Youtube videos, tweets, etc.---\emph{anywhere} on the web, and lets each user independently decide whose assessments they \emph{trust} and want to use to help protect them from misinformation. This tool is an extension of the Trustnet social platform previously offered by Jahanbakhsh et al.~\cite{jahanbakhsh2022leveraging}.
Upon visiting a page, 
the user will see assessments of the content by their trusted sources, if any have contributed one, as shown in Figure~\ref{fig:teaser}. Such assessments can aid the user in deciding whether the content they are viewing is credible. However, since links on social media are often not clicked, even when they are shared~\cite{gabielkov2016social}, showing the assessments of a page only when a user visits the page will not benefit those who do not navigate to the page. To address this issue, the extension also places any available trusted accuracy assessments next to outgoing links on the page the user is reading, as displayed in Figures~\ref{fig:twitter_link} \&~\ref{fig:youtube_link}.


\aptLtoX[graphic=no,type=html]{\begin{figure}[t]
  \centering
  \includegraphics[width=\linewidth]{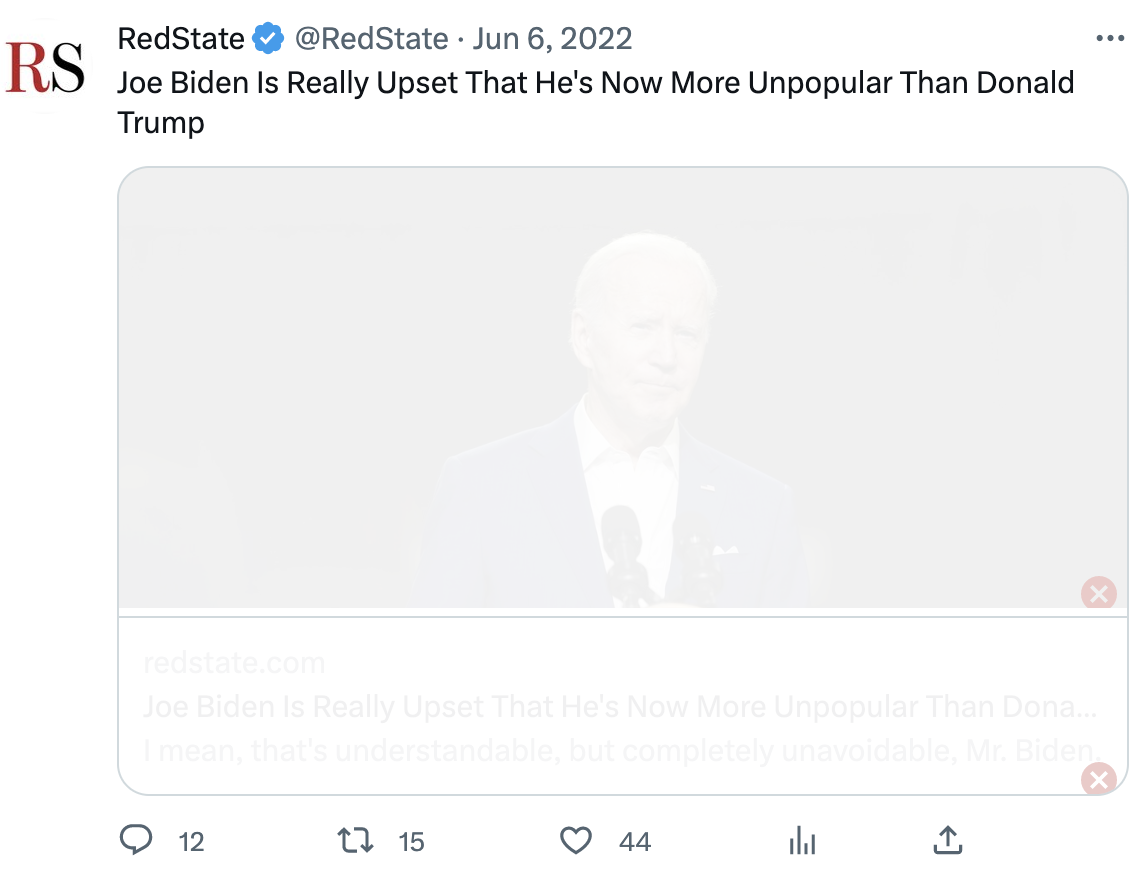}
  \caption{The extension places any available accuracy assessments from the user's trusted or followed sources next to outgoing links on the page the user is visiting. The screenshot shows an example of a red X marking such a link on the Twitter timeline. The link is visually faded because it has been assessed as inaccurate.}
 \label{fig:twitter_link}
 \Description{The image shows a tweet from RedState with a preview of the article that it has tweeted about. The preview is visually faded and appears with 2 cross in a circle icons (one on the image, the other on the title and description), to indicate that the article has been assessed as inaccurate.}
\end{figure}

\begin{figure}
\includegraphics[width=\linewidth]{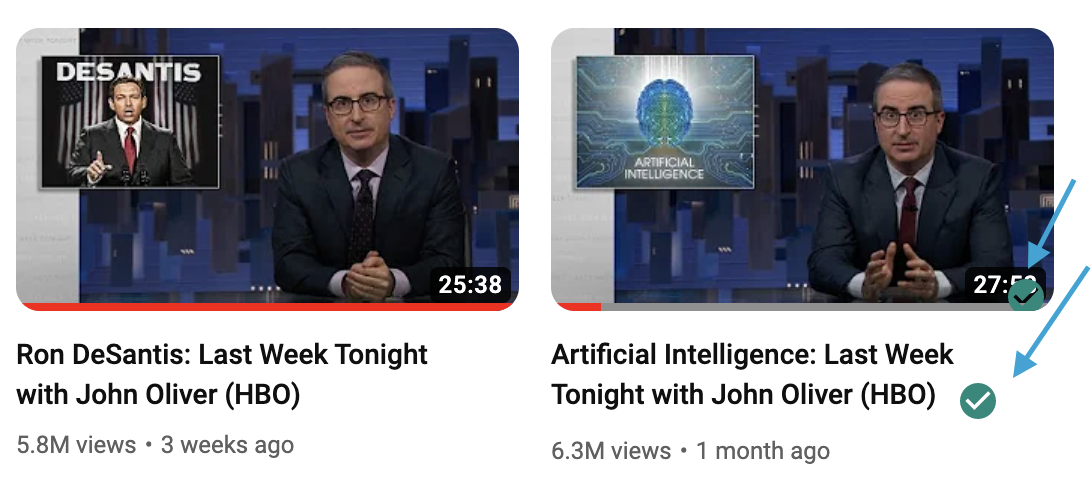}
  \caption{Outgoing links to a Youtube video on the page are displayed with a checkmark icon (marked with blue arrows), indicating that the video has been assessed as accurate by the user's assessors.}
 \label{fig:youtube_link}
 \Description{The image shows previews of 2 Youtube articles, one with a green checkmark in a circle placed on its preview image and title to indicate that it has been assessed as accurate by the user's assessors.}
\end{figure}}{
\begin{figure*}[t]
  \centering
\begin{minipage}{.4\textwidth}
 \centering
  \includegraphics[width=\linewidth]{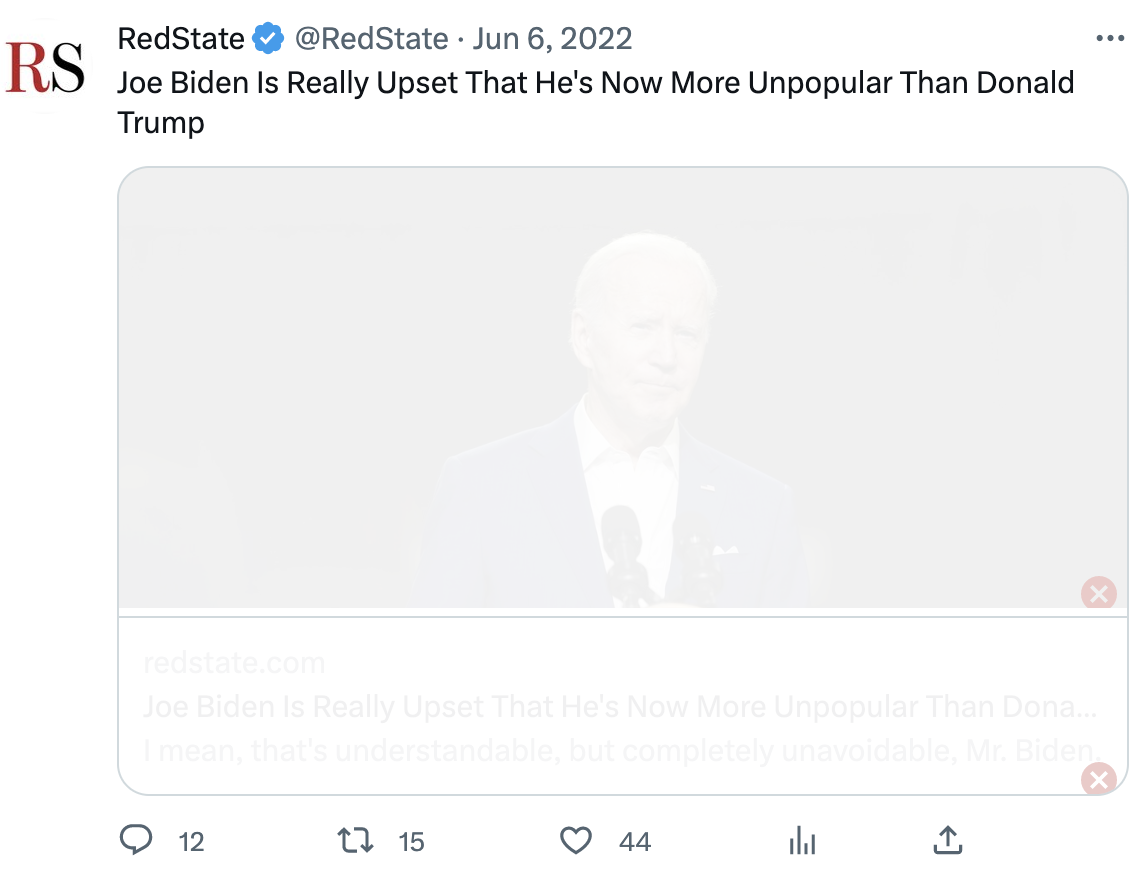}
  \captionof{figure}{The extension places any available accuracy assessments from the user's trusted or followed sources next to outgoing links on the page the user is visiting. The screenshot shows an example of a red X marking such a link on the Twitter timeline. The link is visually faded because it has been assessed as inaccurate.}
 \label{fig:twitter_link}
 \Description{The image shows a tweet from RedState with a preview of the article that it has tweeted about. The preview is visually faded and appears with 2 cross in a circle icons (one on the image, the other on the title and description), to indicate that the article has been assessed as inaccurate.}
\end{minipage}
\qquad
\begin{minipage}{.55\textwidth}
\includegraphics[width=\linewidth]{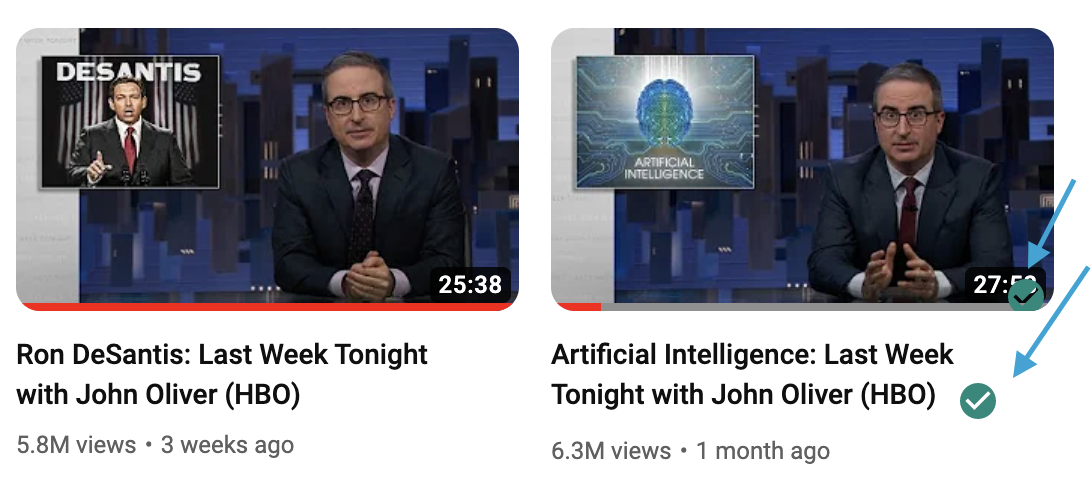}
  \captionof{figure}{Outgoing links to a Youtube video on the page are displayed with a checkmark icon (marked with blue arrows), indicating that the video has been assessed as accurate by the user's assessors.}
 \label{fig:youtube_link}
 \Description{The image shows previews of 2 Youtube articles, one with a green checkmark in a circle placed on its preview image and title to indicate that it has been assessed as accurate by the user's assessors.}

   
 \end{minipage}
\end{figure*}}

To evaluate the potential of such a design and users' perception of it, we conducted a user study of the extension. We recruited 32 users and to use the extension for two weeks. Users were tasked with assessing the accuracy of two pieces of content daily. Users also could, but were not required to, share content with the other users via the extension. The shared content would be presented on each user's feed on the Trustnet platform instance that we asked them to visit on a daily basis. We surveyed our users after the user study to understand their fact-checking needs and desires and their perception of the tool.
The user study serves as a technology probe~\cite{hutchinson2003technology} where users can interact with a tool that empowers them to have an active role in moderating misinformation for themselves and others, without needing to drastically change where and how they consume content. By offering the experience of a new alternative, our technology probe breaks our subjects free of the status quo, encouraging less constrained thinking about their needs and desires for misinformation moderation.

Our user study sets out to answer three main research questions: 

\textbf{RQ1:} What are the facets of users' information needs that a democratized misinformation moderation, made possible through our proposed tool, can address?;

 Our first research question aims to \emph{empirically} demonstrate the potential for democratization of misinformation governance. Investigating the diversity of users' needs after their exposure to the tool helps us understand whether centralized fact-checking can sufficiently cover the breadth of user needs and desires, some of which users may be empowered to articulate after they realize other forms of misinformation moderation may be possible. We focus on multiple aspects of users' needs: how important they think it is to avoid misinformation on various topics, how easy they find it to fact-check content on various topics, whether they want to see assessments from others, and whether they want to contribute assessments for the benefit of others.
 We show that even in our small-scale study, the distribution of the types of content people deem important to assess differs across individuals and from what is typically fact-checked by professional fact-checking initiatives. Moreover, depending on the topic, users wished for including viewpoints and assessments from \address{sources other} than professional fact-checkers alone---e.g., journalists from other countries and immigrants on issues related to foreign affairs. These observations further bolster the argument that there is a need for decentralized approach to content moderation.

\textbf{RQ2:} What criteria for assessing content accuracy do users use when using our browser extension?
    
Examining the rationales that users use in their assessments helps us determine whether regular people can indeed assess content effectively. By extending a taxonomy of such rationales with new categories not previously reported in prior work, we contribute to the literature on user-centered indicators of content credibility~\cite{zhang2018structured, jahanbakhsh2021exploring}. Additionally, these rationales can be built into tools to capture users' assessments of content credibility in structured form.


\textbf{RQ3:} How do users perceive a browser extension for in-situ signaling and assessment of misinformation from their customized sources?
    \begin{itemize}
        \item RQ3a. What do they perceive as the advantages and downsides of the approach?
        \item RQ3b. What are their ideas for improving it and how could they be incentivized to use it?
    \end{itemize}
    
Investigating these questions helps us determine whether users find the tool of value to them and how it can be improved.



The main contributions of our work are: 1) the design of a system for democratized in-place signaling and assessment of misinformation anywhere on the web which allows users to moderate misinformation for themselves with the help of those they trust; 2) a taxonomy of criteria for assessing content credibility not previously reported in prior work emerged from our study, which can be used to extend the credibility indicators framework offered by prior work~\cite{zhang2018structured, jahanbakhsh2021exploring}; and 3) An empirical understanding of how users use and perceive this system.




This tool provides users with the autonomy to moderate and filter misinformation in-situ anywhere on the web for themselves and to additionally help their social circle do so. 
We offer recommendations for extending the design and architecture behind this tool to domains beyond misinformation, and designing a class of tools that further empower users to moderate content for themselves and choose who they want as their moderator.

\section{Related Work}
Throughout the paper, we use the term ``misinformation'' as an umbrella term encompassing the various types of false information that can be found on the web per Zannettou et al., including fabricated stories, propaganda, conspiracy theories, hoaxes, biased or one-sided content, rumors, clickbait, and more~\cite{zannettou2019web}.
Here, we situate our work in the literature related to identifying and dealing with misinformation, tools that allow for personalizing user's web experience, and systems that aid in sensemaking.

\subsection{Identifying and Dealing with Misinformation}
Given the impact that misinformation on social platforms has had on people's lives and even democracy~\cite{argentino2021qanon, abortionMisinfo, bengali2019whatsapp, brennen2020types}, researchers as well as platform operators have been investigating and deploying approaches to detect and counter misinforming content. 
The space of these approaches is broadly contained within these two axes:
\begin{itemize}
    \item Who gets to decide what is misinformation 
    \item and whether the moderation decisions are contained within a single platform/website or applied ubiquitously
\end{itemize}

The arbiters of misinformation can range from a central authority to a body of a trusted few to a democratized process e.g., through majority votes from users to each individual.

\subsubsection{Centralization}
The dominant approach is centralization of the arbitration power---in most cases to the platforms---who decide on both the detection of misinformation as well as the decision on how to deal with the misinforming content. Centralized approaches to misinformation detection include using machine learning models for classifying whether content is accurate~\cite{atanasova2019automatic, qian2021hierarchical, bracsoveanu2021integrating, zeng2021automated}, and leveraging human moderators and third party fact-checking and news organizations~\cite{facebookFactchecking, ananny2018partnership}. Centralized approaches to dealing with misinformation include displaying warning labels next to, down-ranking, reduction (e.g., shadow-banning), or removal of the content or suspension of offender accounts~\cite{facebookVaccine, FacebookTechCrunch, mosseri2016news, FacebookDownRankTechCrunch, chowdhury2021examining, gillespie2022not, myers2018censored}.

A body of work has investigated the effectiveness of these centralized approaches. For instance, Bak-Coleman et al. show that content removal and account banning can be effective at stopping the growth of misinformation~\cite{bak2022combining}. Researchers have also reported that warning labels can increase user discernment of content accuracy or reduce users’ likelihood of sharing misinformation~\cite{yaqub2020effects, bhuiyan2021nudgecred, seo2019trust}. Other studies however, found warning labels have limited or adverse effects~\cite{clayton2020real, gao2018label, pennycook2020implied}.

Nevertheless, the role of platforms as moderators of content and speech is contested by some researches, scholars of law, as well as users who believe the downsides of ceding such control to the platforms can be grave. One such downside is that by assuming the role of a truth arbiter, platforms can limit users' freedom of speech and listening and their autonomy in deciding what content to consume~\cite{jahanbakhsh2022leveraging, grafanaki2018platforms, koltay2021protection, saltz2021encounters}. Related is the concern that even platform-assigned labels or messages can be perceived as truth governance and a threat to freedom of speech~\cite{dillard2005nature, jia2023embedding}. In fact, some users perceive platform labels as judgmental, paternalistic, and against the platform ethos~\cite{saltz2021encounters}. Another concern is that any centralized decision, by platforms or otherwise, to for instance, down-rank or filter misinformation cannot address the needs of every user, as some users want to be aware of what misinforming content their social circle is exposed to, so that they can talk to them about it~\cite{atreja2022will, jahanbakhsh2022leveraging, malhotra2020covid19, rader2015understanding}. Yet another concern is that for these centralized approaches to work, users have to rely on the trustworthiness, goodwill, and competence of the moderators~\cite{mccroskey1999goodwill, lin2016social}; however, some users consider platforms profit-driven and politically biased~\cite{saltz2021encounters, jahanbakhsh2022leveraging}. These concerns are valid given the history of platforms blocking content that arguably did not have potential to harm or content by activists or dissidents in certain autocratic countries~\cite{twitterMastodonBan, FacebookPragerU, instagramModerators, facebookCensorshipNWord, facebookCensorshipActivistsOfColor}.

While the centralized approaches mentioned above are generally contained within a single platform, some exist that have been applied ubiquotiously. These include classifiers to label clickbait or sensationalist headlines based on lexical features that were incorporated into browser extensions that warn users when they view such a headline~\cite{rony2018baitbuster, chakraborty2016stop}.

\subsubsection{A Select Body of a Trusted Few}

There exist approaches that put the power of arbitration in the hands of a trusted few (e.g., fact-checkers or journalists). ClaimBuster is a system of this nature, which checks claims in live discourses, news, and social media posts against a repository of fact-checked claims~\cite{hassan2017toward}.
Another is ConsidertIt, a system for public dialog where participants discuss pros and cons of issues, integrated an on-demand fact-checking into the Living Voters Guide (LVG) deliberation space. Fact-checking was staffed by librarians and any LVG user could request that any pro or con point submitted by other users be fact-checked. The fact-checking information would then be placed in-situ~\cite{kriplean2014integrating}. The interactive fact-checking element in this system is similar to the one in our tool. 
Another example is Videolyzer that enables political bloggers and journalists to assess aspects of quality in videos including accuracy and bias~\cite{diakopoulos2008annotation}. Other systems have also been developed to help the body of the select few monitor and verify potentially critical content shared in social spaces~\cite{melo2019whatsapp}.

While these initiatives are valuable, their effectiveness can be limited by being constrained to a platform and not incorporated into the systems where users encounter content. This issue is countered by tools such as NewsGuard, a browser extension that shows detailed transparency and credibility scores given by journalists to various sources (i.e., websites)~\cite{lapowski2018newsguard}. These scores are placed alongside links that users encounter as they browse the web. A shortcoming of displaying credibility ratings at the granularity of a source is that not all content from an unreliable source is inaccurate; and conversely, not all content from a reliable source is accurate~\cite{jahanbakhsh2022leveraging}. However, for approaches that confine the power to assess credibility to a limited set of individuals, e.g. journalists, scaling the investigation and signalling of the credibility at \emph{content-level} would be impractical. Even at the source-level, such an approach fails to capture the reliability of sources that are not notable enough.

\subsubsection{Democratization}
Moving farther away from centralization of the arbitration power, a body of work has studied how to involve users in the process of content moderation. Most notably, Twitter Community Notes (previously known as Birdwatch) enables users to anonymously write ``notes'' critiquing or adding additional context to a tweet that they find misleading. Other community members can rate notes as helpful. Twitter's bridging algorithm then decides which notes are eventually shown to users by inspecting whether the raters seem to come from diverse perspectives~\cite{diversityTwitterCommunityNotes}. Subreddits and Facebook groups also have moderators from the user body appointed to the role.
In these cases however,  the resulting decision is imposed on all users of the community, regardless of whether they agree with the outcome or the decision making process.

\subsubsection{Individualization}

At the end of the centralization/ decentralization axis lie approaches that do not impose a single assessment for content on all users and instead enable each individual to make more informed decisions about content credibility, e.g., by relying less on intuition and instead, engaging in critical thinking~\cite{pennycook2019lazy, mosleh2021cognitive}. These include nudges that ask users to assess the accuracy of content before sharing it~\cite{jahanbakhsh2021exploring, epstein2021developing}, subtly priming users to have accuracy on top of their mind~\cite{pennycook2021shifting, pennycook2020fighting}, or providing users with checklists based on recommendations from authoritative sources (e.g., the WHO) to guide users in evaluating content reliability for themselves~\cite{heuer2022comparative}. 
Jahanbakhsh et al. proposed a set of design affordances, that they argue platforms should adopt, which democratize the power to not only determine what content is misinforming, but also decide what to do about such content. These affordances include enabling users to: 1) assess content as part of the data model 2) specify whose assessments they consider trustworthy, and 3) filter misinformation from their feed as assessed by their trusted sources.
These affordances are intended to empower users in their practices around fighting misinformation that the authors uncover through a study--- users already ask those who they trust within their social circle to verify the information that they encounter. And they also provide fact-checking information to their friends and family~\cite{jahanbakhsh2022leveraging}. Indeed, users are more receptive to correcting information from their friends than strangers~\cite{hannak2014get, margolin2018political}.
However, platforms do not support these user practices and sometimes even undermine them. 

Jahanbakhsh et al. provided these user affordances through a prototype social media platform that they designed~\cite{jahanbakhsh2022leveraging}. However, it is unrealistic to expect that the myriad of websites and platforms where users consume content will adopt these affordances. Therefore, in this work, we explore using a browser extension that acts as an overlay on the web, and allows users to assess the accuracy of any content that they encounter across platforms, as well as see the assessments of their trusted sources on the content. Our design deals with operationalizing trusted assessments ubiquitously and in-situ including where they should be placed considering the particulars of user's content consumption practices on the web; e.g., the fact that many users skim headlines and links to content, but click on them only sporadically~\cite{boczkowski2017incidental}. In addition, our architecture broadens the coverage of content that can be assessed beyond articles.

There do exist other tools that both are individualized and whose arbitration decisions are applied ubiquitously. One is the Reheadline browser extension that enables users to suggest headlines they deem better for news articles and specify whose suggested headlines they want to see on the web~\cite{jahanbakhsh2022our}.
Another is Dispute Finder, a browser extension that highlights text snippets making disputed claims and shows articles arguing for and against the claim. The 
disputed claims in the system are restricted to those in certain fact-checking websites such as Snopes and Politifact. With centralized fact-checking often lagging behind the spread of false claims~\cite{starbird2014rumors}, the effectiveness of this tool can be limited when the user encounters content that is not old enough to have been fact-checked by fact-checkers of the two platforms. Dispute Finder is individualized as it is the user who specifies the sources of the articles a user wishes to see for or against the disputed claims, although users are restricted to websites that meet the Wikipedia criteria for being reliable~\cite{ennals2010highlighting}.


\subsection{Personalizing Users' Web Experience}

Our browser extension belongs to the class of tools that personalize  users' web experience across platforms. A classic example of such a tool was WebWatcher, which followed the user's actions on the web, learned the user's interests, and highlighted or added certain hyperlinks on the page for the user to follow~\cite{joachims1995webwatcher, joachims1997webwatcher}.
Within this class, web annotation tools exist that enable users to add content to webpages where they do not have authoring permission. One such example is Hypothesis, a browser extension that allows users to highlight and annotate text and for future visitors on the page to view and interact with the annotations. Hypothesis users can make their annotations visible to the public or only within a private group of users~\cite{Hypothesis, kalir2019open}. NB is a similar social annotation tool for classroom settings~\cite{zyto2012successful}. Other tools have also existed that enable users to add comments on a page that are not necessarily tied
to a particular piece of content on the page. Examples include Google Sidewiki and Eyebrowse~\cite{GoogleSidewiki,zhang2016opportunities}.

Our extension differs from these tools in two ways. One is that the assessments, i.e., the annotations or metadata, that are submitted via the extension are structured. The structure makes it possible to differentiate assessments arguing for the accuracy of the content and the counter arguments as well as the questions about the accuracy of the content. The differentiation makes it possible to show each of these categories differently in the UI. For helping the user determine content credibility as is the purpose in this domain, it is important to surface arguments as well as counter-arguments in a salient fashion. The other difference in this tool is the fine-grained control that the user has in determining their sources of assessments, i.e., their trusted sources after the model proposed in~\cite{jahanbakhsh2022leveraging}.

\subsection{Systems that Aid in Sensemaking}

By allowing users to share assessments with each other, and to contextualize their assessments in-situ, our browser extension is related to a class of systems that empower and aid users in the process of sensemaking. The sensemaking brought about by our tool is in fact distributed---i.e., it leverages other users' work without those users directly collaborating with each other~\cite{fisher2012distributed}. Research has shown that empowering users to scaffold on each others’ learning can increase the depth of their sensemaking~\cite{kittur2014standing}.
Of the other work that similar to our case, aid users in making sense of information, Goyal et al. provided a sensemaking translucence interface for crime solving, which included a hypothesis window to promote idea exchange and a suspect visualization for automatic feedback on suspects discussed in the hypothesis window~\cite{goyal2016effects}. The role of visualization in improving sensemaking has been explored in other work as well, including for representing data links or analysis provenance ~\cite{goyal2013effects, li2020crowdtrace}. Other work has reported on the effectiveness of creating shared representations between domain experts and the crowd at increasing sensemaking~\cite{venkatagiri2019groundtruth}.
Close to our scenario, Kuznetov et al. designed a browser extension with the goal of reducing the friction of collecting information while remaining in the flow of the sensemaking process. The extension allows users to access and organize external content on the web in-situ while preserving the provenance of and the context around the content~\cite{kuznetsov2022fuse}.

\section{The System}
To build our system, we extended the open source Trustnet platform presented in~\cite{jahanbakhsh2022leveraging}. The reason we used Trustnet was because it already provides an API for managing the data models that our envisioned extension would use. In this section, we provide a summary of the Trustnet platform and describe the design of the extension that we built.

\subsection{The Trustnet Platform}
Trustnet is an open source prototype social content sharing platform developed by Jahanbakhsh et al~\cite{jahanbakhsh2022leveraging}. On this platform, users can post content or import content from other websites using their URL, \emph{follow} other users or news media (all generally referred to as \emph{sources}) to see the content they post on their newsfeed, provide assessment of content, i.e., mark content as accurate or inaccurate and provide their reasoning, \emph{trust} other users, and see the assessments of those they follow and trust on the content appearing in their newsfeed. The platform also offers a filter which a user can use to filter the posts in their newsfeed based on how the posts have been assessed by the user's trusted sources. Users can also inquire about the accuracy of posts, and optionally elaborate on their question. A user's question about content accuracy is relayed to their trusted sources by default; although the user can choose to specify from whom they would like assessments. Users can ask their question about content accuracy anonymously. On this platform, who a user trusts is kept private to the user. Follow relationships on the other hand, are public. 

We extended the API of Trustnet's backend to serve the browser extension that we developed. In addition, we updated Trustnet's client to work with the extended backend. Using the extended system, users can sign up for an account on the Trustnet platform, and specify which other users they trust or want to follow, and then use the same account for logging into and working with the extension.

\subsection{The Trustnet Extension}
\begin{figure*}[t]
  \centering

   \includegraphics[width=\linewidth]{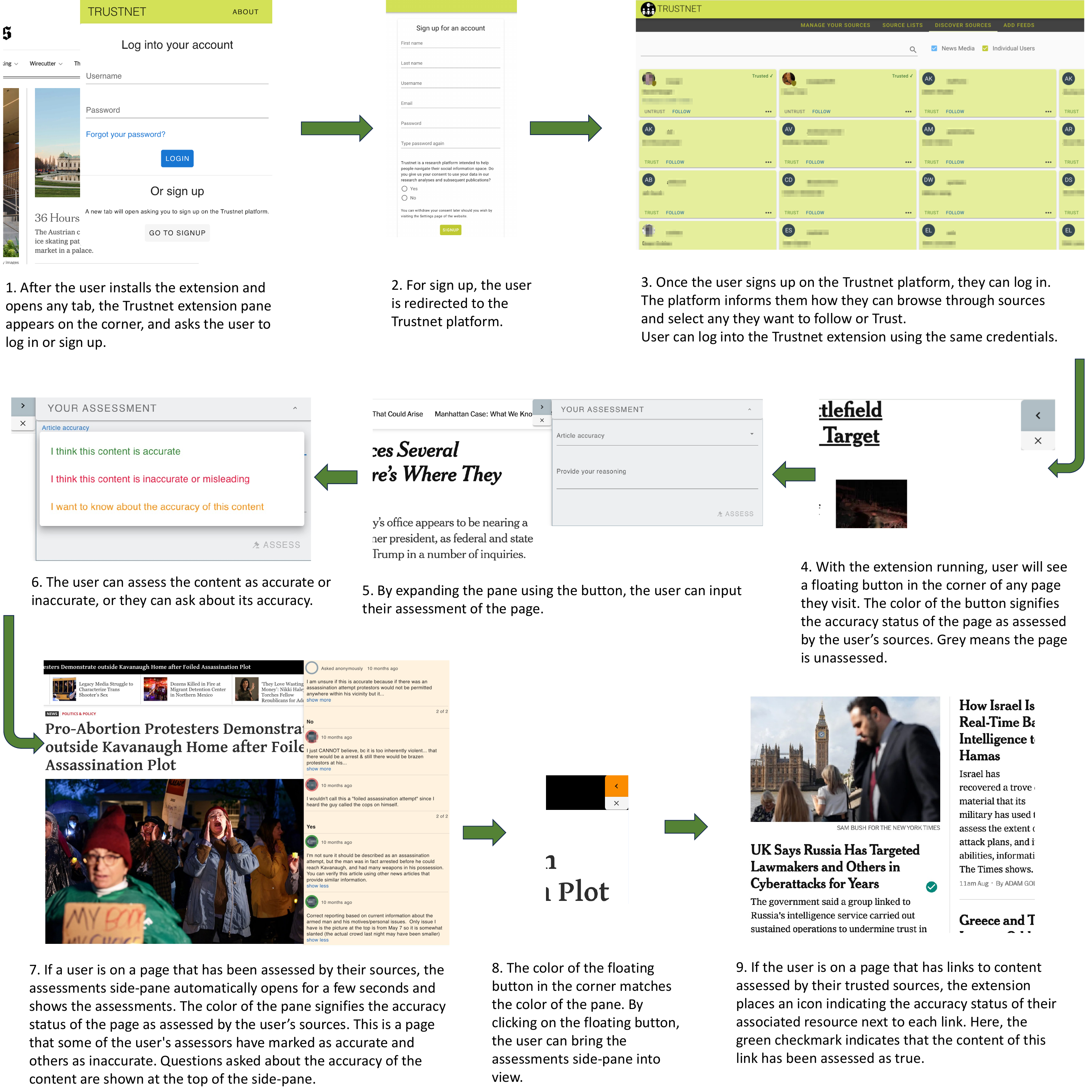}
  \captionof{figure}{The flow of user interaction with the Trustnet extension, from signing up to assessing and seeing assessments from others.}
 \label{fig:user-flow}
 \Description{The image shows 9 interactions with the Trustnet system: 1. After the user installs the extension, on the first tab that they open, the Trustnet extension pane appears on the corner, and asks the user to log in or sign up.
 2. For sign up, the user is redirected to the Trustnet platform.
3. Once the user signs up on the Trustnet platform, they can log in. The platform informs them how they can browse through sources and select any they want to follow or Trust.
User can log into the Trustnet extension using the same credentials.
4. With the extension running, user will see a floating button in the corner of any page they visit. The color of the button signifies the accuracy status of the page as assessed by the user’s sources. Grey means the page is unassessed.
5. By expanding the pane using the button, the user can input their assessment of the page.
6. The user can assess the content as accurate or inaccurate, or they can ask about its accuracy. 
7. If a user is on a page that has been assessed by their sources, the assessments side-pane automatically opens for a few seconds and shows the assessments. The color of the pane signifies the accuracy status of the page as assessed by the user’s sources. This is a page that some of the user's assessors have marked as accurate and others as inaccurate. Questions asked about the accuracy of the content are shown at the top of the side-pane.
8. The color of the floating button in the corner matches the color of the pane.
9. If the user is on a page that has links to content assessed by their trusted sources, the extension places an icon indicating the accuracy status of their associated resource next to each link. Here, the green checkmark indicates that the content of this link has been assessed as true.
}
\end{figure*}

The design decisions pertaining to the Trustnet extension~\footnote{Code available at https://github.com/farnazj/Trustnet-Extension\newline
Extension on the Chrome web store:\newline https://chromewebstore.google.com/detail/trustnet/nphapibbiamgbhamgmfgdeiiekddoejo} were made through many rounds of discussion within the research team as well as pilot tests with a broader group of researchers who volunteered their time as alpha testers. Here we describe the design as well as the rationales behind the various decisions. The flow of the user interaction with the system is shown in Figure~\ref{fig:user-flow}.

When a user is on a web page, the extension checks the URL (and its variations\footnote{Depending on how the page is served, the same resource can be accessed with slightly different URLs, for instance with or without \textsf{index.html} or \textsf{/} concatenated to the end, over both secure and plain HTTP, or using geolocation IP services to redirect requests to different domain names such as BBC.co.uk to BBC.com if the requester IP is in the U.S.}) to see whether its associated content has been assessed by those the user trusts or follows. If there are such assessments for the page, a pane that contains the assessments pops open on the side of the page automatically. The pane stays open for a few seconds and then is minimized to a floating button on the top right corner of the viewport to avoid obstructing the user's view. The pane and the floating button color is determined by the assessed accuracy of the content, as we will later describe. Clicking on the floating button re-opens the pane. The reason we made the decision to show the pane automatically when assessments were present, as opposed to letting the user notice the colored floating button, was because we expected that with a small user base, users would only occasionally encounter pages which their sources had assessed and we wanted to prevent ``change blindness''. Change blindness is a phenomenon where changes in a visual scene go unnoticed because of visual inattention. The degree of change blindness is higher when the observer does not expect a change~\cite{rensink2005change}, e.g., because change happens infrequently. We deemed it important to successfully deliver the signal that assessments are present on a page. With a wider adoption and a higher likelihood of assessments being present on various pages, users could be given the option to have the pane minimized from the beginning and open automatically in certain situations, for instance, when the page has been evaluated as inaccurate.
For ease of discernment, the color of the pane as well as the floating button signals the accuracy status of the content: green for accurate, red for inaccurate (Figure~\ref{fig:teaser}), and orange for split opinion---where some of the user's assessors have marked the content as accurate and others as inaccurate (See Figure~\ref{fig:user-flow}).

The accuracy status of the page (and hence the pane's color) is determined by: 1. the assessment that the user has submitted for the page, if any. 2. In case the user has not assessed the page, the assessments from those that the user trusts if any exist. 3. If there are no such assessments, the assessments of those the user follows. A page is determined as accurate if \emph{all} the assessments of the relevant assessors (i.e., the user, or those that the user trusts, or those that the user follows in absence of trusted assessments) have evaluated the page as accurate. The same is true for inaccurate. If there is disagreement among the relevant assessors, the page's accuracy is determined as ``split opinion". If the page has no assessments from either the user or those that the user trusts or follows, the pane contains no assessments and its color is grey.

We chose the green for accurate and red for inaccurate color scheme to be externally consistent with fact-checking platforms such as Politifact and Snopes. We selected the (yellow-)orange color for split opinion because it is halfway between green and red on the color wheel. Another design that we considered for the split opinion status was for the color of the pane to be determined as a distance between red and green based on the proportion of assessments marking the content as accurate vs inaccurate. However, in cases when a minority assessment was outnumbered by assessments arguing the opposite, it was possible that the color would be so close to either red or green that it would be indistinguishable. We determined it was important to signal even a lone outlier disagreement amidst one-sided assessments, and that perhaps such contrary viewpoint can point the user to the specific trusted assessor who might help guide them out of an echo-chamber. Therefore, we decided against using a color gradient.

The same pane also shows inquiries about the accuracy of the content from those that either trust the user who is visiting the page or have specifically asked that their question be relayed to the user. The user can use the same pane to submit their own assessment of the page, update their previous assessment, or ask about the accuracy of the content. They can also use the pane to share the content into the Trustnet platform after they have assessed it or asked about its accuracy. By sharing the content, other users who follow the sharer will see the content in their Trustnet feed along with the sharer's assessment of it. To help users expand their network, the pane informs the user if there exist assessments on the page by someone the user does not follow or trust. The pane shows up to 10 such sources who are trusted by the most number of users platform-wide. The user can choose to follow any of these sources if they wish to see their assessments on the page. Users can expand, collapse, or close the extension pane.

An advantage of using URLs to bind assessments to and find assessments of content is that they are universal. Any resource on the web including news articles, social media posts (e.g., Facebook posts and tweets) and even their comments, Youtube videos, etc. has a unique URL and therefore the user can see its assessments by their social circle when navigating to the content's URL. However, many of these resources are usually embedded into a feed (e.g., social media posts) or are presented as links on an aggregator page (e.g., the front/home page of a news website). Prior work has reported that links, for instance on social media, are often not clicked, even when they are shared~\cite{gabielkov2016social}. Therefore, showing the assessments of a page when a user visits the page will not benefit the multitude of those users who do not navigate to the page. To address this issue, the extension also checks the links on the page that the user is presently reading, and places an icon indicating the accuracy status of their associated resource next to each (if there exists assessments for the resource from the user's network) by slightly modifying the rendered page (see Figure~\ref{fig:youtube_link}). The extension further places a question mark next to the links about the accuracy of whose content other users have inquired. We additionally made the design decision to reduce the salience of links assessed as inaccurate by making them look faded (somewhat transparent but still visible), as seen in Figures~\ref{fig:twitter_link} \&~\ref{fig:youtube_link}. The motivation for reducing the visual salience was to reduce the ``illusory truth'' effect, in which a single prior exposure to a headline increases subsequent perceptions of its accuracy~\cite{pennycook2018prior}. By attempting to bias eye movements (a proxy for overt attention) through reducing the visual salience of content assessed as inaccurate, we aimed to make it easier for users to ignore links that are misinforming~\cite{itti2001computational}.  In effect, we tried to introduce a design friction as a disincentive---a small hurdle that users would need to overcome to read the content~\cite{ohm2018desirable}. Nevertheless, users can still consume the content if they wish to as the content is still where it would otherwise be and is visible.


\subsubsection{Dealing with Link Complications}
A complication of finding assessments of outgoing links on a page is that these links cannot be used as they are found on the rendered page to check whether they have associated assessments in the Trustnet system's backend.
The reason is that many links go through (sometimes multiple) redirections before finally redirecting to the link from which the resource can be fetched. For instance, a link shared on X (formerly known as Twitter) is automatically processed and shortened to another with the format of \code{https://t.co/xxx}. Links on other websites may also first redirect to an intermediary URL so that the website logs outgoing links (i.e., which link a user clicks to leave the page). 

To address the issue of redirections, when the extension encounters links on a page, it recursively follows each~\footnote{In order to protect against broken looping links, there is a maximum depth after which the extension abandons the pursuit.} until it encounters the final link that contains the resource. The reason it is the client rather than the server that follows the redirects is three-fold: First, the number of links a user encounters on any one page can be large, and if all the links encountered by all the users are sent to the server so that it can follow their trail of redirects, the process can incur a substantial delay as there is a limit on the number of parallel requests the server can send and therefore, it would need to batch the requests sequentially. Second, it is conceivable that many of the requests across users are to the same websites and if a large number of requests are made to these websites by one or a few servers, the servers can get rate-limited.
Third, the information to which the client has access allows it to visit some links that the server cannot, for instance, those that are behind paywalls. Nevertheless, the client cannot follow all the links as there are occasions when upon following the redirects, the browser encounters a Cross-Origin Resource Sharing (CORS) issue. The CORS issue typically occurs when a web application running in one domain (the ``origin") tries to make an HTTP request to a resource located on a different domain. If the server hosting the resource does not explicitly allow the requesting domain to access it, the browser will block the request.
This for instance, happens on Facebook or Twitter where the encountered links are shortened links from the social media domain but the target links belong to other domains which enforce the CORS policy~\footnote{For example: https://t.co/CUnCxRezGn (origin on Twitter) -> https://www.cnn.com/politics/live-news/election-live-updates-11-07-23/ (resource on CNN.com)}. Because CORS is only enforced in the browser, these links are sent to the server for following.

It is likely that the links which have been encountered and whose trails have been followed will be encountered again by the same or other users. Therefore, to boost future performance, after following the links and fetching their target links, the client creates a mapping of the (cleaned\footnote{The extension first cleans the links found on the page, for instance by adding missing protocol or hostname.}) original to the target links and sends it to the server. The server then caches this mapping. When visiting a page, the extension first sends the encountered links on the page to the server and asks whether it holds a mapping for any of them, as fetching the targets is faster this way. With enough people using the extension, the load of following the link redirects on any one client becomes lower, as it will be more probable that the server already knows a mapping for at least some of the requested links. The server discards the mappings that have not been requested for some time. 

Most redirects are either HTTP redirects (returning HTTP code 300) or HTML redirects (where the URL to follow can be found in a \textsf{meta} element). Another redirection method is doing so in the JavaScript running on the page which is harder to detect. We wanted the extension to be general-purpose and to work on as many platforms as possible. During development, we tested the extension on numerous news and content sharing websites where we anticipated users would want to post assessments. One where we found JavaScript redirection was Google News. We made specific accommodations to fetch redirect URLs from returned HTML of Google News URLs.

Many URLs have query parameters appended to them that are usually optional key value pairs customizing the query. For instance, clicking on an article link on Facebook will append a query parameter of the form \code{?fbclid=xxx} to the article's URL, which is used for tracking purposes. Before requesting assessments for URLs or requesting mappings of them from the server, the extension strips the URL of these query parameters because they do not play a part in identifying the resource and their inclusion would result in a false miss in the cache. However, we found a few websites where we anticipated users may want to use the extension and which unconventionally use query parameters to distinguish a resource. These websites include Youtube (where video URLs are of the form \code{https://www.youtube.com/watch?v=xxx}), Facebook photos, videos, and comments (but not for instance, Facebook posts), and Hacker News posts (where post URLs are of the form \code{https://news.ycombinator.com/item?id=xxx}). We made specific arrangements so that the extension treats query parameters from these websites as a resource.

The set of links on a page can change even as the URL of the page is constant, for instance, because more content is loaded as a result of an infinite scroll. The extension watches the page for changes and reacts by looking for new links for which to fetch assessments.

The extension additionally has an Options page where the user can specify their blacklisted domains, i.e., where they do not want the extention to run.

\section{User Study of the Tool}
We recruited participants to work with the Trustnet browser extension and the Trustnet platform.
We advertised the study on a behavioral research platform managed by our institution with a pool of diverse participants. We asked those who were interested to fill out a survey which in addition to describing the study and providing the consent form, asked about the news sources participants consumed and the subreddits, Facebook groups, and Twitter accounts that they frequented or followed which discussed news. We collected this information to test the extension on these websites or subcommunities prior to the onset of the study and resolve any potential issues. In the survey, we also asked prospective users to refer others in their network who might be interested in participating because the tool can best be leveraged among members of a social circle with pre-existing trust relationships. However, we were unsuccessful in recruiting those that were referred. The survey additionally asked for demographic information, adopted from prior studies on misinformation~\cite{jahanbakhsh2021exploring, jahanbakhsh2022leveraging}.

For onboarding, we asked that users watch a short video tutorial explaining the systems and their features, sign up for an account on the Trustnet platform (which would enable users to log into the Trustnet extension as well), and add the extension to their Chrome browser through the Chrome Web Store. To bootstrap the study, we configured every user to initially follow all other users of the study upon signup, so that they would be able to see each other's assessments. Users could later choose to unfollow one another.
We asked that users perform the following tasks each day during the two-week period of the user study:
\begin{itemize}
    \item Assess the accuracy of two pieces of content per day. These could be news articles, tweets, Youtube videos, or any other piece of content that the participant wanted. We encouraged our participants to assess as many more pieces of content as they liked.
    \item Check out the Trustnet feed to see what the users of the study share every day. The reason was because through simply using the extension, with such a small user base and varied interests, it was unlikely that users would serendipitously run into each other's assessments as they were browsing the web. We encouraged, but not required users to share content to the Trustnet feed through the Trustnet extension. The reason we did not require sharing was because a valid use of the system---a feature missing in existing social platforms---is assessing content as misinformation for the benefit of whoever comes across the content, but not wanting to further increase its visibility. 
\end{itemize}

At the study's conclusion, participants completed two surveys asking about their experience with assessing, seeing assessments from others and their perceptions of the tool. The survey included questions such as ``To what extent did you like being able to assess content'' and ``To what extent did you like to see the assessments of other users?'' with 5-item likert scale responses as well as free-text elaboration. To better understand users' thought process in assessing content, the survey showed each user examples of content they had assessed as inaccurate and asked them to explain the characteristics of content they assessed as (in)accurate. The questions on the experience of assessing included if users believed their assessments helped others and what content they had difficulty assessing if any, among others. The questions on the experience of receiving assessments included how well-thought-out users found others users' assessments and their accounts of disagreements with others. Some of the survey questions were inspired by prior work on democratized misinformation moderation~\cite{jahanbakhsh2022leveraging}.

To compensate participants, we randomly selected 4 via a raffle and awarded each with a \$100 gift card.
After analyzing the results, we observed that users had assessed a wide range of content on different topics, not necessarily limited to political news. However, their responses in the post-study survey centered around assessment of political content, possibly because a few questions had primed them to have politics on top of their mind. Therefore, we sent our participants another follow-up survey to further understand their thoughts on how useful they would find assessments on different types of content and their current as well as desired fact-checking practices around different topics. This survey included questions such as ``How important is it for you, if at all, to avoid inaccurate information related to each of these topics? (likert scale)'', ``Who (if anyone) would you like to see assessments from on each of these topics? Please elaborate'', and ``to what extent do you think you can contribute assessments that can help others on each of these topics''.
Participants were each compensated with a \$10 gift card for completing this follow-up survey.
The questionnaires are included in the Supplementary Materials. The study was approved by our Institutional Review Board.

\subsection{Participants}
A total of 32 users participated in the user study, of whom 25 submitted assessments for 3 or more days. View the full participation plot in Appendix Section~\ref{appendix:participation-plot}. Out of all the participants, 21 completed the post-study survey and 25 completed the follow-up survey asking about assessments on different topics. 66\% were female and 25\% were male. 28\% identified as Democratic, 19\% as Republican, and 53\% as Independent. The median age was 26 (ranging from 19 to 67), the highest degree achieved Bachelor's degree (ranging from high school diploma to Doctoral degree or equivalent), and the median income \$60,000 -  \$69,000 (ranging from less than \$10,000 to \$150,000 or more).

\section{Results}
\label{section:results}
 For categorizing the themes surfaced in users' qualitative data such as their assessments or answers to the surveys, one member of the research team made multiple passes over the data and used open coding to assign codes to the idea units and axial coding to consolidate the themes~\cite{strauss1998basics}. These codes were refined through subsequent passes over the data by constant comparison and theoretical sampling (i.e., building a temporary theory and then collecting more data to test that theory), resulting in the emergence of more generalizable concepts. Then through selective coding, the researcher focused more narrowly on codes of topical interest considering the question at hand~\cite{muller2010grounded}, e.g., whether there is potential for democratization or personalization of misinformation moderation. During the analysis stage, we followed up with some participants via email to ask them to clarify some of their responses.
 As described, our use of grounded theory was interpretive and aimed to yield concepts and themes; therefore, agreement or measures of inter-rater reliability were not appropriate~\cite{mcdonald2019reliability}.

\subsection{The Breadth of User Needs Supports the Case for Democratized Misinformation Moderation (RQ1)}

\subsubsection{Users Differ with Each Other and with Fact-checkers on the Importance of the Accuracy of Various Topics.} 
\begin{figure*}[!t]
 \centering
 \includegraphics[width=\linewidth]{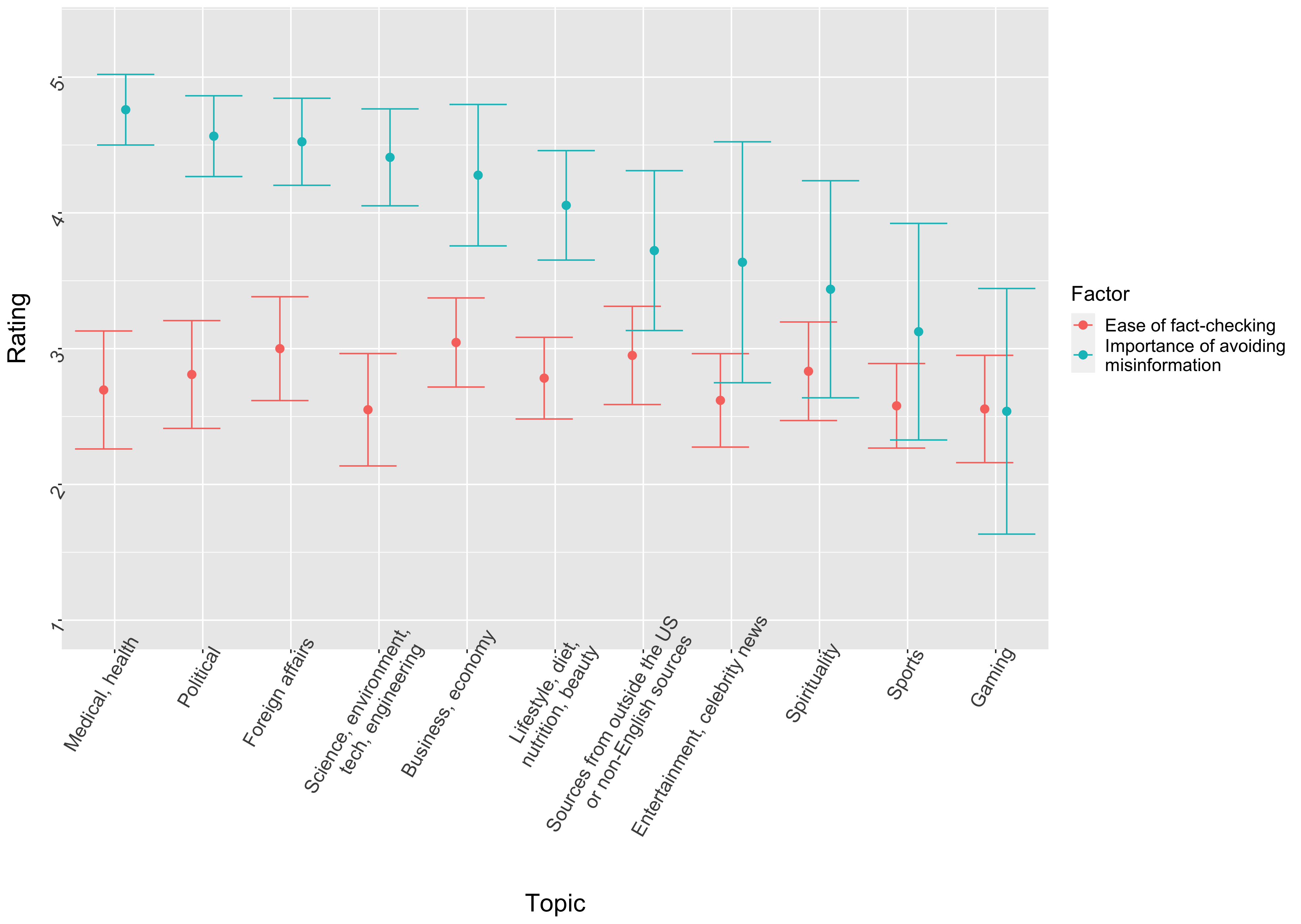}
 \captionof{figure}{The extent to which participants found it important to avoid misinformation on various topics, and how easy they found it to fact-check information related to the topics they reported at least a little important to avoid misinformation on (rating >= 2), on a scale of 1-5. Each user answered these questions only on those topics that they reported they consume.}
 \label{fig:importance-of-misinfo-avoidance-fact-checking-ease}
 \Description{The figure shows means and standard errors around the means for participants' responses to the importance of avoiding misinformation on various topics as well as the ease of fact-checking content on those topics. The topics include: medical \& health, political, foreign affairs, science \& environment \& tech \& engineering, business \& economy, lifestyle \& diet \& nutrition \& beauty, sources from outside the US or non-English sources, entertainment \& celebrity news, spirituality, sports, and gaming. The order of the means for the average importance of avoiding misinformation from highest to lowest is the order of the topics mentioned previously.}
\end{figure*}

Responses in the follow-up survey revealed that users not only differ in the topics that they consume, but also to what extent they find it important to avoid misinformation related to each topic. For instance, while some said that the accuracy of any content matters to them, others considered certain topics of less relevance or importance to them; some believed certain topics are inherently non-factual or controversial, or that it is impossible to find unbiased news related to certain topics.
Figure~\ref{fig:importance-of-misinfo-avoidance-fact-checking-ease} shows the perceived importance of avoiding misinformation related to different topics averaged among our participants. The accuracy of health-related and political content was at least of moderate importance to most users. However, even in our small user sample, there was some variation in how important they found the accuracy of other topics compared to the topics that are often scrutinized by fact-checkers. For instance, one participant deemed it more important for them to avoid misinformation on spiritual content compared to political or scientific content. Another considered the accuracy of content on lifestyle, diet, nutrition, and beauty more important than that on foreign affairs. The status-quo of centralized fact-checking constrained to a limited set of professionals is unlikely to match with the breadth of the types of content for which different users want assessments.

Similarly, the ease of fact-checking content related to each topic, and therefore the need for the involvement of external fact-checkers, varied across users. Some participants explained that they fact-check by reading articles on a topic from different sources (N=6); others relied on their trusted news sources to have done their research (N=3); some said certain topics are harder to fact-check because they are rarely presented objectively or that they do not have objective standards by which they can be assessed (N=4), they are complex and the user does not have the required background knowledge to fully understand them or that the knowledge in the field is constantly evolving (N=5), or that it takes a great deal of effort to assess the accuracy of an article on the topic along various criteria such as missing information and statistics (N=4).
Figure~\ref{fig:importance-of-misinfo-avoidance-fact-checking-ease} shows how easy participants on average find it to fact-check content on various topics.  

\begin{quote}
    \textit{``The more dense topics such as medical questions or in-depth political issues require a lot of foreknowledge or quick education to understand what is being discussed before you can fact-check or make an opinion on something.'' (P25)}
\end{quote}

\subsubsection{There is a desire and potential for including others in providing assessments, than professional fact-checkers alone.}
The viability of an ecosystem that involves users beyond professional fact-checkers in misinformation moderation is dependent on two factors: 1) Whether users would want to view assessments from such assessors, and 2) whether they can contribute assessments that can help others.

To investigate the first aspect, in the follow-up survey, we asked users from whom they want to see assessments on content related to each topic in an extension like the Trustnet extension that allows them to see assessments from those they follow and trust. We have summarized the responses per topic in Table~\ref{tab:desired-assessors}. For all of the topics, users wanted to see the assessments of experts and professionals as well as other users. However the notion of what qualifications the experts or what characteristics the users should have varied across participants. These responses corroborated prior work reporting that users want to and do seek the judgment of a variety of different sources they consider trustworthy to determine whether content is credible~\cite{jahanbakhsh2022leveraging}.

What we additionally uncovered was that depending on the topic, participants wished to also view assessments from a broader set of individuals. For instance, on political content, participants liked to see assessments from users with certain or diverse demographics or political leanings or those users who have been affected by certain policies. On issues related to foreign affairs, they wished for the assessments of users or journalists from other countries and immigrants.

\begin{table*}[!t]
\begin{footnotesize}
\setlength{\aboverulesep}{0pt}
\setlength{\belowrulesep}{0pt}
\setlength{\extrarowheight}{.35ex}
\newcolumntype{g}{>{\columncolor{gray!16}}c}
\setlength{\tabcolsep}{6pt}
\caption{Participants' responses to who they would like to see assessments from on content related to various topics. Examples of assessors belonging to various categories that participants cited are shown in parentheses.}
\Description{Fully described in the text.}
\label{tab:desired-assessors}
\centering
\begin{tabular}{p{2cm}|p{3cm}|p{2.5cm} |p{3.2cm}|p{1.1cm}|p{0.9cm} |p{2cm}}
  \toprule
 \cellcolor{gray!16} {Topic} & \cellcolor{gray!16}{Professionals \& experts} & \cellcolor{gray!16}{Academics \& researchers}  &\cellcolor{gray!16} {Users} & \cellcolor{gray!16}{Journalists} & \cellcolor{gray!16}{Friends \& family} &\cellcolor{gray!16} {Other} \\
  \midrule

     \cellcolor{gray!16} Political & N=11 ( political analysts, politicians, previous office holders, judges) & N=5 (Political scientists, historians, etc.) & N=9 (those from diverse demographics or political leanings)  & N=5  & N=2 & N=4 (certain media outlets from all sides, political activists) 
     \\
      
\hline

         \cellcolor{gray!16} Medical, health   & N=17 (medical specialists) & N=9 & N=2 (patients) & N=1 & N=1 & 
         \\

\hline

     \cellcolor{gray!16} Science, environment, tech, engineering & N=9 (those in tech, designers, engineers) & N=9 (the researchers who conduct the studies being written about, peers in the field) &	N=4  &	N=2	& N=1 &  
     \\
\hline

     \cellcolor{gray!16} Business, economy & N=9 (business analysts, CEOs) & N=4 \newline	(economists) 
 & N=5 (small business owners)	& N=3 &	N=1  & 
     \\

   \hline

     \cellcolor{gray!16} Foreign affairs & N=6 (diplomats, ambassadors, government workers) & N=3 & N=7 (locals, people on both sides of issues) & N=8 (foreign correspondents & N=1 
     \\
\hline

     \cellcolor{gray!16} Lifestyle, diet, nutrition, beauty & N=6 (medical professionals, life gurus, etc.) & N=3 & N=6 (women, those interested in weight loss or skincare) & & N=1 & N=1 (influencers)
  \\
    
 \hline
     \cellcolor{gray!16} Sources outside the US or non-English sources & N=2 & N=3 & N=4 (immigrants, international audience, those who speak the language fluently) & N=5 & & N=1 (firsthand sources)
     \\

 \hline

     \cellcolor{gray!16} Sports & N=10 (athletes, coaches, retired sportspeople) & N=1 & N=2 (fans) & N=3 & N=1 
    \\

\hline

     \cellcolor{gray!16} Gaming & N=6 (gamers, e-sports professionals) & & N=2 & & &
    \\

\cline{1-7}

  \cellcolor{gray!16} Entertainment, celebrity news & N=3 & & N=2 & N=4 & N=2 &
    \\

\hline
     \cellcolor{gray!16} Spirituality & N=6 (pastors and spiritual leaders from certain backgrounds or specializing in certain denominations ) & N=2 (theologians) & N=5 (religious users or those with similar moral backgrounds as the user) & & & \\

  \bottomrule
\end{tabular}
\end{footnotesize}

\end{table*}

To investigate the second aspect, we asked our participants to indicate on which topics they believe they can contribute assessments that can help others. In Section~\ref{section:criteria}, we will further shed light on this aspect by reporting on the types of rationales they used in their assessments.
Two outcomes that would pose challenges for the viability of such an ecosystem would be if users believed themselves incapable of assessing or if on the contrary, they were ready to profess their opinion about credibility of content without consideration. What we observed however, was that the extent to which participants perceived their own assessments on various topics helpful varied considerably and depended on whether they were an expert on, interested in, or well-read on a certain topic, their confidence in their knowledge, whether they could trust themselves to be unbiased, and whether they considered themselves capable of finding accurate and balanced information to pass their findings to others. Therefore, it was interesting that users exhibited discernment about when their assessments would contribute meaningfully, as demonstrated in this quote:

\begin{quote}
\textit{``I know more about certain topics because I work in that area or because I am interested and consume more about those topics. I feel that my assessments would be more useful in those areas rather than in areas I have no experience or interest in.'' (P20)}
\end{quote}

\subsubsection{Users Assessed Content from Diverse Sources }
  Our participants submitted a total of 641 assessments and asked 39 questions about the accuracy of various pieces of content. 
To understand what type of content users would want to assess if given the opportunity, we investigated the sources to which the content belonged. The sources of content that participants assessed or inquired about using the Trustnet extension included various local, national, and global news, hyper-partisan, or misinformation websites (e.g., BBC, CBS58 Milwaukee, National Review, Info Wars, Common Dreams), news aggregators (e.g., AllSides, Yahoo News),
non-English news sources (e.g., El País), government websites (e.g., CDC), academic journals (e.g., National Library of Medicine), publications of various universities (e.g., NYU Law and University of Bath UK), posts on content sharing websites (e.g., Twitter, Youtube, and Hacker News), and specialty news and blogs (e.g., related to the economy, sports, entertainment and celebrity gossip, diet and nutrition, home improvement, etc.). Source of other content assessed included archives and personal content (e.g., email). 

The tweets and Hacker News posts that participants had assessed
had embedded links to outside news articles. The assessed Youtube videos were created by accounts that had between ~200K and 10M (in the case of major media organizations) subscribers. The content that these videos discussed were diverse including some that touted certain misleading information and conspiracy theories, others that explained various scientific and historical events, some that disclosed the unethical practices of various corporations or the inauthenticity of certain claims such as those of psychics and mediums, and, because it was making news at the time, a number that discussed Johnny Depp vs Amber Heard trials, for instance, inviting body language and behavior experts to ``reveal what she really thought''.

\subsection{Users' Criteria for Assessing Content (RQ2)}
\label{section:criteria}
We investigated users' assessments to understand what kind of criteria and rationales they had used for evaluating accuracy. To analyze the themes in participants' rationales, we first consulted prior work that presented criteria for assessing content accuracy: the taxonomy of users' rationales for assessing news headlines~\cite{jahanbakhsh2021exploring}, the context and content credibility indicators developed by the Credibility Coalition in an attempt to define a shared language for classifying misinformation and credibility~\cite{zhang2018structured}, and problems with news headlines (e.g., clickbait, sensationalism, etc.) that users corrected when given a tool that enabled them to do so~\cite{jahanbakhsh2022our}. To analyze the rationales that emerged in the assessments (and questions---since some questions also contained rationales for why the content may or may not be accurate) of our users, we first partitioned their responses into idea units, resulting in 732 idea units, each being a coherent unit of thought. Of these, a total of 204 idea units did not include rationales, e.g., because they were questions about the accuracy of the content, or they simply summarized the article. We then assigned codes from this body of related work to the rest of the idea units when the codes were appropriate, noting which rationales could not be captured by any of the proposed codes in the related work. The criteria presented in prior work and discussed by our users included the trustworthiness of the source, presenting evidence that corroborates the claim, presentation of all viewpoints, soundness of inference or logic, citation of studies, organizations, or experts, representative citations, title representativeness, among other factors. We then used grounded theory as described in Section~\ref{section:results} to build concepts that could encompass the user rationales that were not assigned labels. Some of these emergent concepts were generalizations of, and therefore encompassed, some of the criteria in the prior work. When reporting the number of citations of rationales based on these more general concepts, the count includes instances that were originally coded using the criteria from the earlier work. To test inter-rater reliability for this part of the results, another coder was trained on the criteria from the developed codebook as well as those in~\cite{jahanbakhsh2021exploring, zhang2018structured} that the criteria in the codebook did not encompass. A sample of the idea units (about 12\%) were labeled by the coder. Both the training and test samples were selected such that they
covered all of the categories. Cohen's Kappa was 0.63, indicating substantial agreement~\cite{landis1977measurement}.

Here, we elaborate on some of the criteria that are related to those reported in prior work. We do not elaborate on the rest of the criteria because the definitions accompanied by the examples convey the meaning well. We have summarized all criteria along with representative quotes in Table~\ref{tab:criteria}. These can be used as criteria to extend the list of credibility indicators for content and context developed in~\cite{zhang2018structured}.

\subsubsection{Citations are credible.} Two indicators in~\cite{zhang2018structured} that are closely related to this category include citation of organization and studies and quotes from experts (in the field), as they can add context or support to the article. Our users however, found content credibly not only through researcher and expert citations but also from citations by others, including victim's family, eyewitnesses of various events, cancer patients (e.g., to attest to certain symptoms), and others. Therefore, the notion of who is a credible citation varies by topic.

\subsubsection{The details or underlying data are disclosed adequately.} This category is closely related to the rationale reported in~\cite{jahanbakhsh2021exploring} about why people \emph{believe} news claims---that evidence presented in the article corroborates the claim. In~\cite{jahanbakhsh2021exploring}, there was no counterpart to this rationale for why people would \emph{disbelieve} a news claim. Our users however, when discussing the data or details presented in an article, sometimes reported inadequate or even more than adequate disclosure of details as a criteria that detracts from article credibility.

\subsubsection{The Content Discusses a Scientifically Reproducible Process or a Claim that Can Be Verified.}
This category is related to a rationale reported in~\cite{jahanbakhsh2021exploring} for why people \emph{disbelieve} claims---that the claim references something that is impossible to prove, e.g., using a metric that cannot be measured. In~\cite{jahanbakhsh2021exploring}, there was no counterpart to this rationale for why people would \emph{believe} news claims. In the assessments that our users submitted, we observed that because a claim could be verified (e.g., through a scientific process or by cross-checking the evidence), it made users believe the claim as accurate.
In addition, one use case of this rationale that we observed, and which was not reported in~\cite{jahanbakhsh2021exploring}, was that at the moment when the article was published or being read, the general information or evidence at hand was or was not enough to deduce whether the article was accurate. Or, while the claim in the article was accurate at that moment, the situation was still evolving and future developments might render the claim inaccurate, or vice versa.

\subsubsection{The Content Distinguishes Between Facts and Opinion.}
One of the criteria that our users used for assessing content credibility was whether the content was intended to be a factual piece as opposed to for instance, op-ed or anecdote, and if so, whether the piece exhibited any editorialization. Even if the content was editorialized, users considered author's transparency about separating the facts and the opinions as a signal of content credibility. This category is a generalization of a rationale reported in~\cite{jahanbakhsh2021exploring} for why users \emph{disbelieve} news claims---that the claim appears motivated or biased.



\begin{table*}[!t]
\setlength{\aboverulesep}{0pt}
\setlength{\belowrulesep}{0pt}
\setlength{\extrarowheight}{.35ex}
\newcolumntype{g}{>{\columncolor{gray!16}}c}
\setlength{\tabcolsep}{6pt}
\caption{The criteria for content accuracy cited by the users of the extension which were not reported in prior work on indicators of content credibility~\cite{zhang2018structured, jahanbakhsh2021exploring}. These criteria can be used to extend the list of credibility indicators.}
\Description{Fully described in the text.}
\label{tab:criteria}
\centering
\begin{tabular}{p{3cm}|p{6.5cm}|p{7cm}}
  \toprule
  \cellcolor{gray!16}  {\textbf{Criterion}} & \cellcolor{gray!16}{\textbf{Yes}} & \cellcolor{gray!16}{\textbf{No}} \\

%

 
 
 

  \midrule

     \cellcolor{gray!8} The content distinguishes between facts and opinion. (N=116) &  
    \textit{``It's a good theory; and it's presented as such: a theory. Nothing for sure! So that makes it accurately presented.'' (P14)} 
     & \textit{``Presents fact and personal opinion in the same setting without clarifying appropriately''(P24)}
     \\

  \hline

    \cellcolor{gray!8} The citations are credible. (N=93) & \textit{``This article provides references and published articles/data.'' (P17)}
    & \textit{``… the news source uses an unverified; anonymous Twitter account as its primary source.'' (P1)}
    \\

\hline

    \cellcolor{gray!8} The details or underlying data are disclosed adequately (not less or more than needed). (N=71) & \textit{`Provided quantitative data about the number and location of grocery stores w/in majority and non-majority Black neighborhoods.'' (P7)}
    \vspace{0.15cm}\newline \textit{``I think the inclusion of surveys; statistics; and examples of what makes ads invasive and their prevalence on the internet through this article proves how detrimental they can be to a common user's experience on the internet.'' (P10)} 
    & \textit{``The writing style feels wrong. There are too many details that should have been withheld; and that gives me reason to think it might be sensationalized'' (P33)}
    \vspace{0.15cm}\newline \textit{``The article states over 2000 pre-existing brain scans of people with and without anorexia were used. There are many missing important factors that the article author did not address- 1. how many of the scans belonged to people with anorexia (proportions of data); 2. age groups looked at; 3. gender; 4. length of illness; or length of remission; 5. any other illnesses or pre-existing conditions\newline Appropriate sourcing is used here and the author uses good quotes. I just wish they would have gone a bit more in depth about the actual data; rather than expecting readers to believe the conclusion right away.'' (P26)}
    \\
    
\hline
    \cellcolor{gray!8} The content discusses a scientifically reproducible process or a claim that can be verified. (N=19) & \textit{``Explains a process that is scientifically reproducible.'' (P32)}
    & \textit{``Not entirely accurate. Reporting of a few unattributed quotes that businessfolks in China are against the govt's zero covid19 policy. There is no way of verifying the veracity of this...'' (P4)}
    \\
    
\hline

    \cellcolor{gray!8} The investigation appears to be of high quality. (N=17) & \textit{``Well-researched; numerous sources including the original professor/director; explanations of both sides of the issues involved in lithium-sulfur battery production'' (P27)}
    & \textit{``This article is basically a series of tweets. Reading only a paragraph in will make it apparent that the title is absolute clickbait. There isn't a lot of substance here. This was a missed opportunity to report more about the 260 employees that got layed [sic] off.'' (P26)}
    \\

\hline

    \cellcolor{gray!8} The relevant disclaimers are reported and are salient. (N=8) & \textit{``This article does a good job of vetting the source and study (not peer-reviewed and only 1 study author). It also mentions the author's own; self-admitted weaknesses to the study. Overall this was a balanced take on an unconventional study (may be more accurate to call the study an analysis instead).'' (P26)}
    & \textit{``The caveats named in the story should have been placed higher in the content. This is unfortunately too common among viral health stories.'' (P1)}
    \\
      
\hline

    \cellcolor{gray!8} The content answers the main question it poses coherently and stays on topic. (N=5) & --- & \textit{``The focus of the article is a question that the content does not answer or give any clarity to.'' (P1)}
    \\

\hline
    \cellcolor{gray!8} The visual artifacts (esp the lede image) are appropriate and do not have a biased undertone. (N=4) & \textit{``good use of graphs''} (P24) & \textit{``I could already tell it was biased given the fact that it posted a picture of a protest instead of the actual pregnancy center that got torn down'' (P11)}
    \\

\hline
    \cellcolor{gray!8} Fabricating claims in the piece would not benefit anyone. (N=3) & \textit{`There is nothing at stake for misinformation here. :)'' (P32)}
    & \textit{``The study was done by a group who actively promotes LGBTQ rights. Of COURSE it's going to be a flawed study!'' (P33)}
    \\

\bottomrule

    
\end{tabular}%
\end{table*}


While our users mostly used the tool to assess content accuracy, in some cases, they put it to other use cases which points to opportunities for enabling individuals to annotate content beyond assessment of factuality. For instance, they used the tool to explain that an article was not factual, but satirical or an opinion piece. Other use cases included expressing opinions or sentiments---that a particular story should not have been published (for instance because it distresses the subjects the story is about or the story is not news worthy) or conversely, that the story is insightful, important, or positive, or to express their (dis)agreement with the message of an article. 



\subsection{Participants' Perceptions of the Tool and the Affordances It Provided (RQ3)}

\subsubsection{Perceived Utility of the Tool and Its Advatanges and Downsides (RQ3a)}

Figures~\ref{fig:like-assessing} and~\ref{fig:like-seeing-assessments-from-others} show the extent to which participants liked being able to assess content and seeing the assessments of other users respectively. We investigated their free-text responses to understand their reasons. The reasons they liked being able to assess included that it helped them think about the news in a more analytical way or gauge their trust in a source (N=5), and that they liked being interactive with the news content they consumed and the ability to call out content they found biased or misleading (N=4). The cited downsides of assessing were that it took extra time and effort (N=4) and that sometimes they find it hard to assess a piece of content e.g., when there is no obvious cause to think the article might be inaccurate or worrying that one is biased in one's assessment (N=3).
A frequently cited reason why participants liked seeing assessments from others included because users can get to know other people's perspectives and standards for evaluating content credibility or that it was fun (N=11):

\begin{quote}
\textit{``It was really interesting to see what people thought of different articles, and especially to read the reasons WHY they thought that.  As opposed to people just blindly taking a side, I liked to see people actually articulate and think through their assessments. It was especially interesting to read when people had different views as to whether content was trustworthy or not.'' (P14)}
\end{quote}

Other reasons included that by considering whether they agree or disagree with their assessments the user can think more critically about content (N=2), and assessments from others, especially trusted sources, help the user decide whether they should trust an article (N=2). Some participants however, disliked democratized assessments because they believed people are biased or incapable of assessing (N=2).

\aptLtoX[graphic=no,type=html]{\begin{figure}
 \centering
 \includegraphics[width=\linewidth]{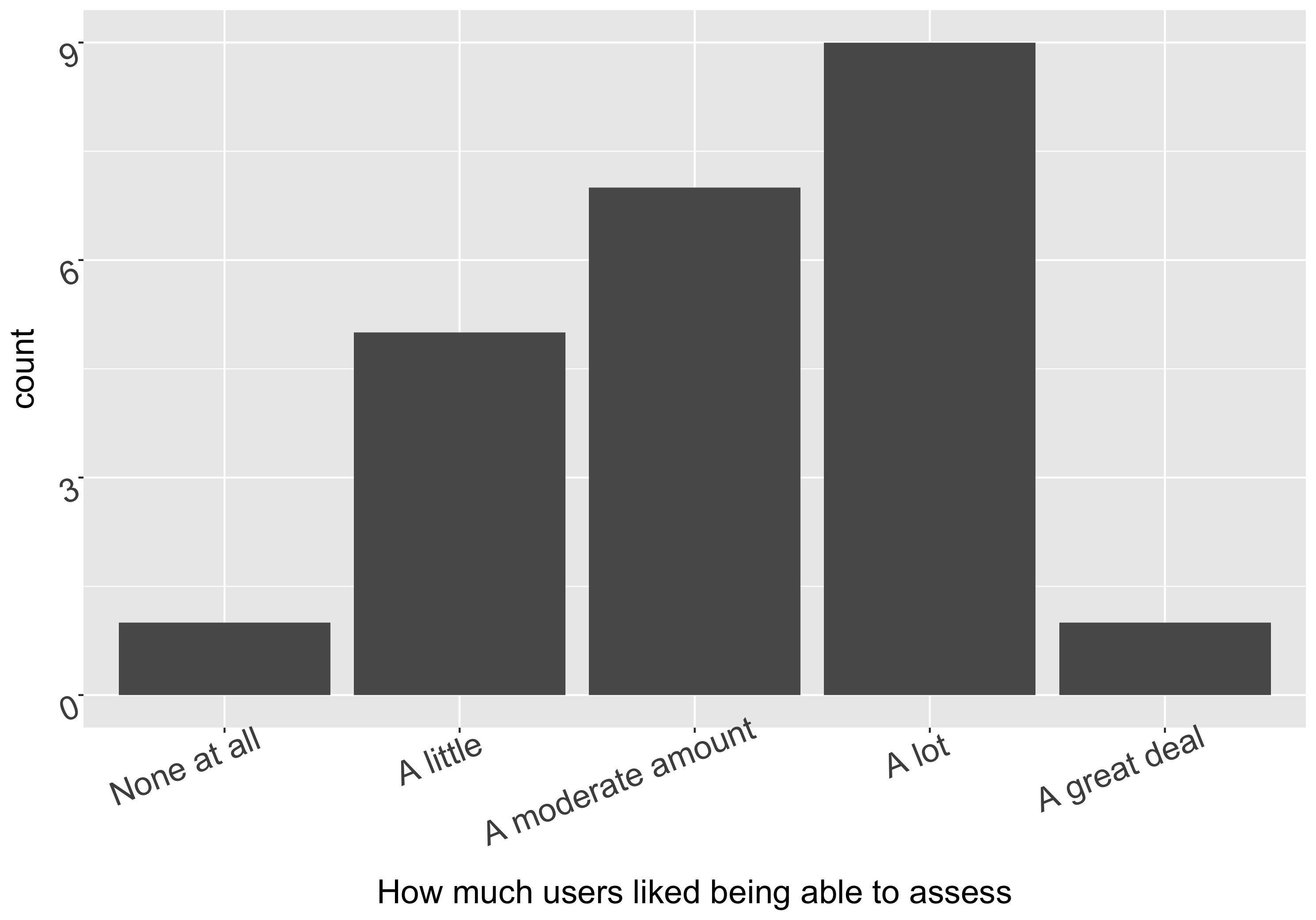}
 \captionof{figure}{The extent to which users reported that they like the ability to assess content in the post-study survey.}
 \label{fig:like-assessing}
 \Description{A bar plot of how much users liked being able to assess content on a 5 point likert scale from ``none at all'' to ``a great deal''. Out of the 23 responses, only 1 has indicated ``none at all'' and 17 have said that they like the ability at least ``a moderate amount''.}
\end{figure}
\begin{figure}
 \centering
 \includegraphics[width=\linewidth]{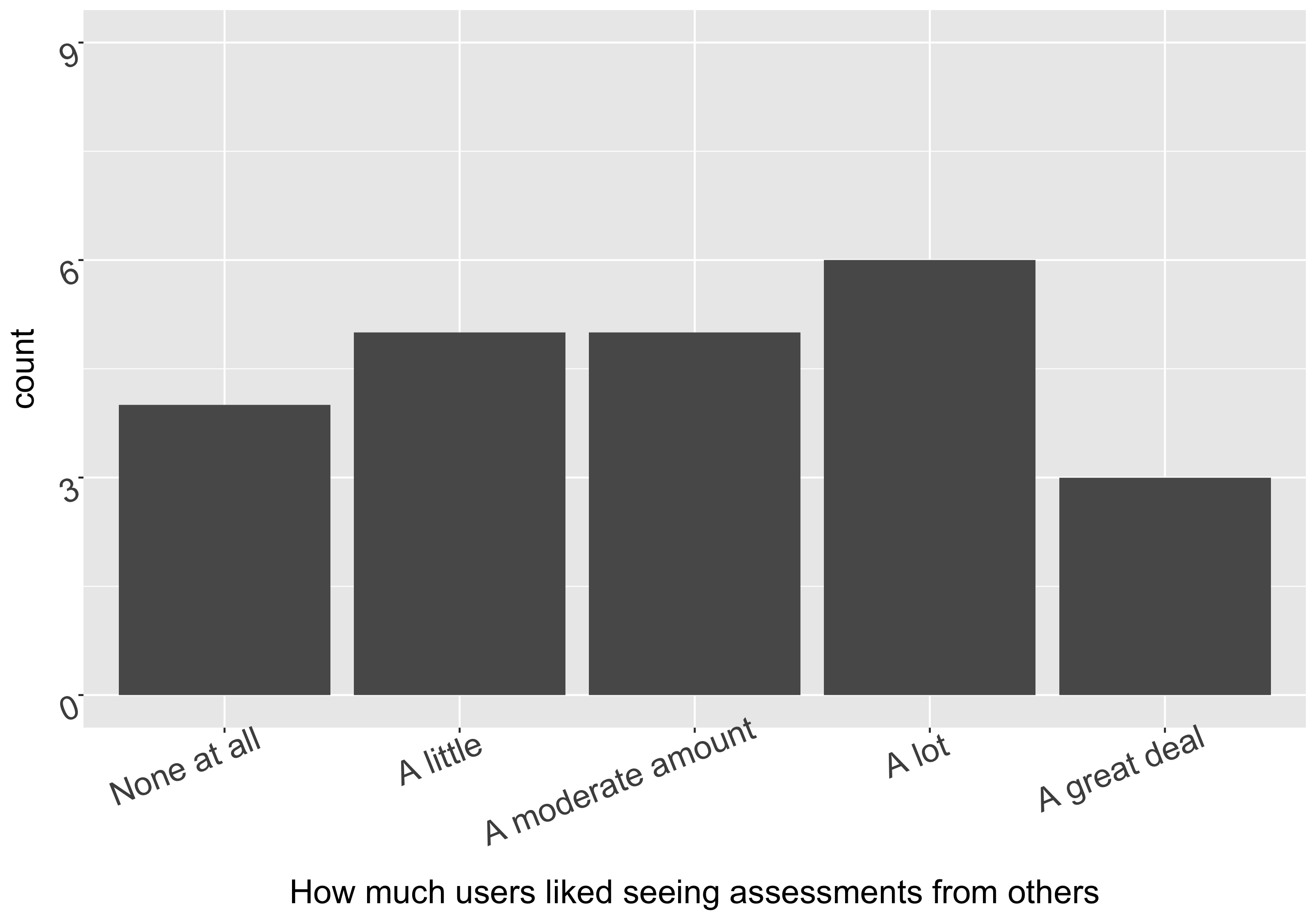}
 \captionof{figure}{The extent to which users reported that they like seeing assessments from others in the post-study survey.}
 \label{fig:like-seeing-assessments-from-others}
  \Description{A bar plot of how much users liked seeing assessments from others on a 5 point likert scale from ``none at all'' to ``a great deal''. The distribution is almost uniform.}
\end{figure}}{
\begin{figure}[!t]
\centering
\begin{minipage}{.47\textwidth}
 \centering
 \includegraphics[width=\linewidth]{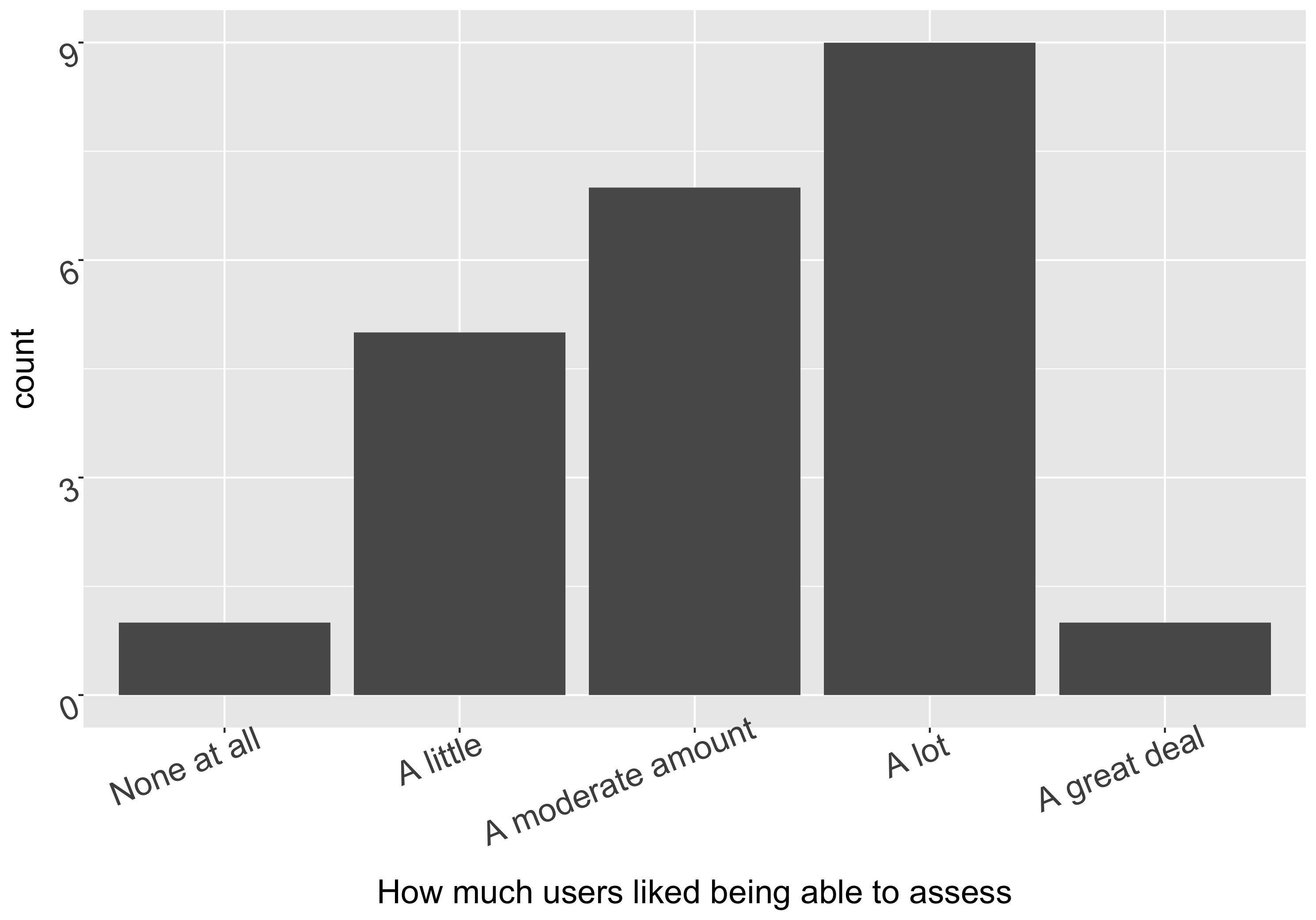}
 \captionof{figure}{The extent to which users reported that they like the ability to assess content in the post-study survey.}
 \label{fig:like-assessing}
 \Description{A bar plot of how much users liked being able to assess content on a 5 point likert scale from ``none at all'' to ``a great deal''. Out of the 23 responses, only 1 has indicated ``none at all'' and 17 have said that they like the ability at least ``a moderate amount''.}
\end{minipage}\qquad
\begin{minipage}{.47\textwidth}
 \centering
 \includegraphics[width=\linewidth]{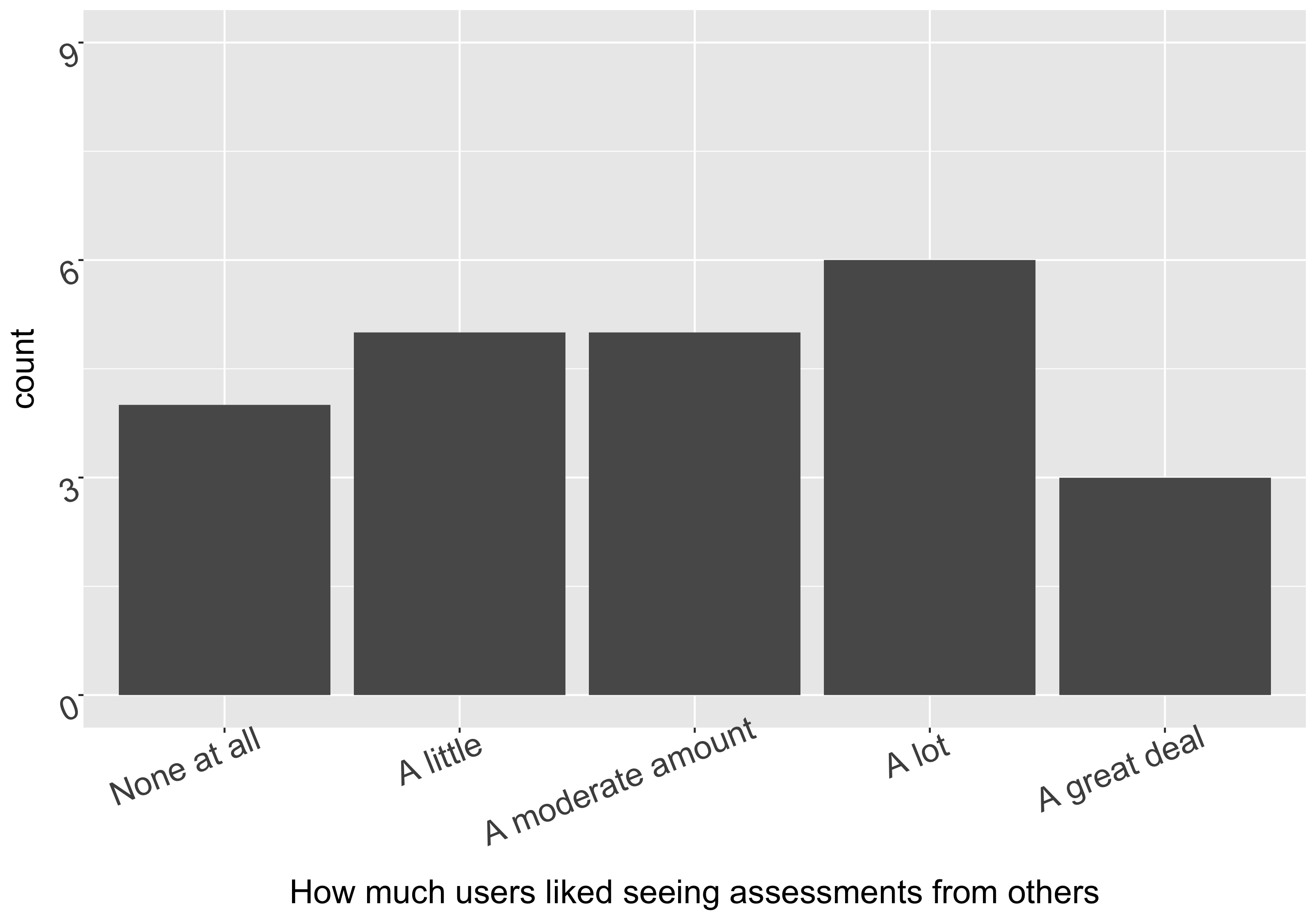}
 \captionof{figure}{The extent to which users reported that they like seeing assessments from others in the post-study survey.}
 \label{fig:like-seeing-assessments-from-others}
  \Description{A bar plot of how much users liked seeing assessments from others on a 5 point likert scale from ``none at all'' to ``a great deal''. The distribution is almost uniform.}
\end{minipage}
\end{figure}}

Figure~\ref{fig:perceived-helpfulness-round-1} shows participants' post-study responses to how much they believe a tool like the Trustnet extension would be useful to them if adopted by many people ($\bar{X}=2.57, s=0.79$). 
In the follow-up survey, we asked participants a similar question of to what extent they find it useful to see the assessments of content right on the page. 
Responses to this question are shown in Figure~\ref{fig:perceived-usefulness-round-2} ($\bar{X}=2.84, s=1.07$).

Many of the rationales (N=14) people cited for why they (dis)liked seeing in-situ assessments were in alignment with reasons why they (dis)liked assessing or seeing assessments from others. Other pros that users discussed were that it in-situ assessments would save others the work of fact-checking that they may not have time for (N=2) and they provide helpful feedback to journalists (N=1):

\textit{``Everyone who has looked into certain topics in-depth should have a right to voice their opinion to others and I think this is a good way of doing so online. It can help improve journalism as they are getting more direct feedback/commenting on their sites instead of comments directed at them but on other social media.'' (P25)}  

 Additional cons that they discussed included that they want to think for themselves, unassisted by anyone (N=4), a sentiment that has been reported in prior work as well~\cite{jahanbakhsh2023AI}, and a worry that users may become accustomed to taking all content with a green checkmark next to it as credible, and not fact-check content for themselves (N=1).

\aptLtoX[graphic=no,type=html]{\begin{figure}
 \centering
 \includegraphics[width=\linewidth]{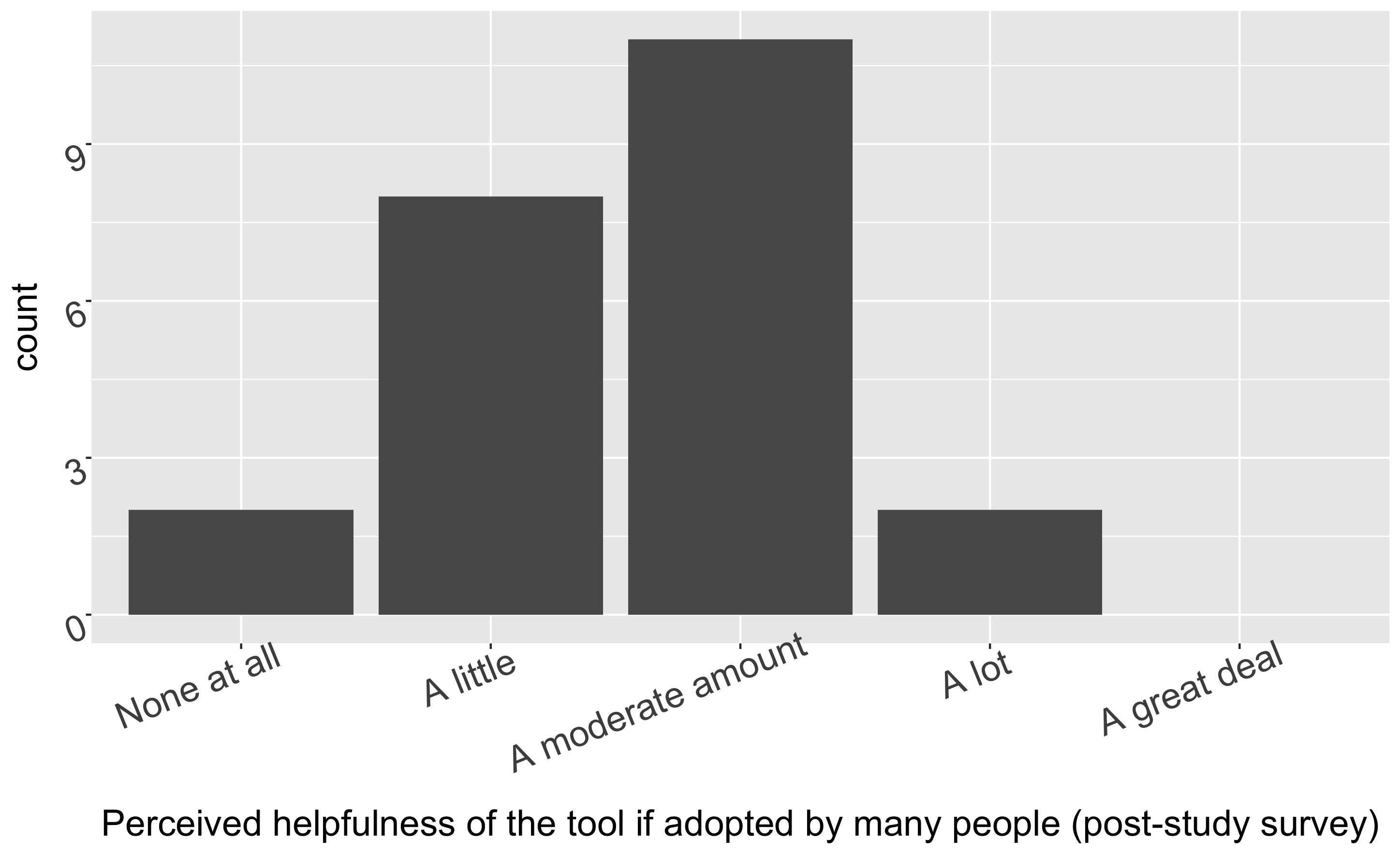}
 \captionof{figure}{The extent to which participants reported the tool would be useful if adopted by many people on a 1-5 likert scale, in the post-study survey.}
 \label{fig:perceived-helpfulness-round-1}
 \Description{A bar plot of how much participants believed the tool would be useful to them if adopted by many people, in the post-study survey. All but 2 of the respondents said the tool would be helpful to them at least a little. 13 out of the 23 respondents found the tool helpful at least a moderate amount.}
\end{figure}
\begin{figure}
 \centering
 \includegraphics[width=\linewidth]{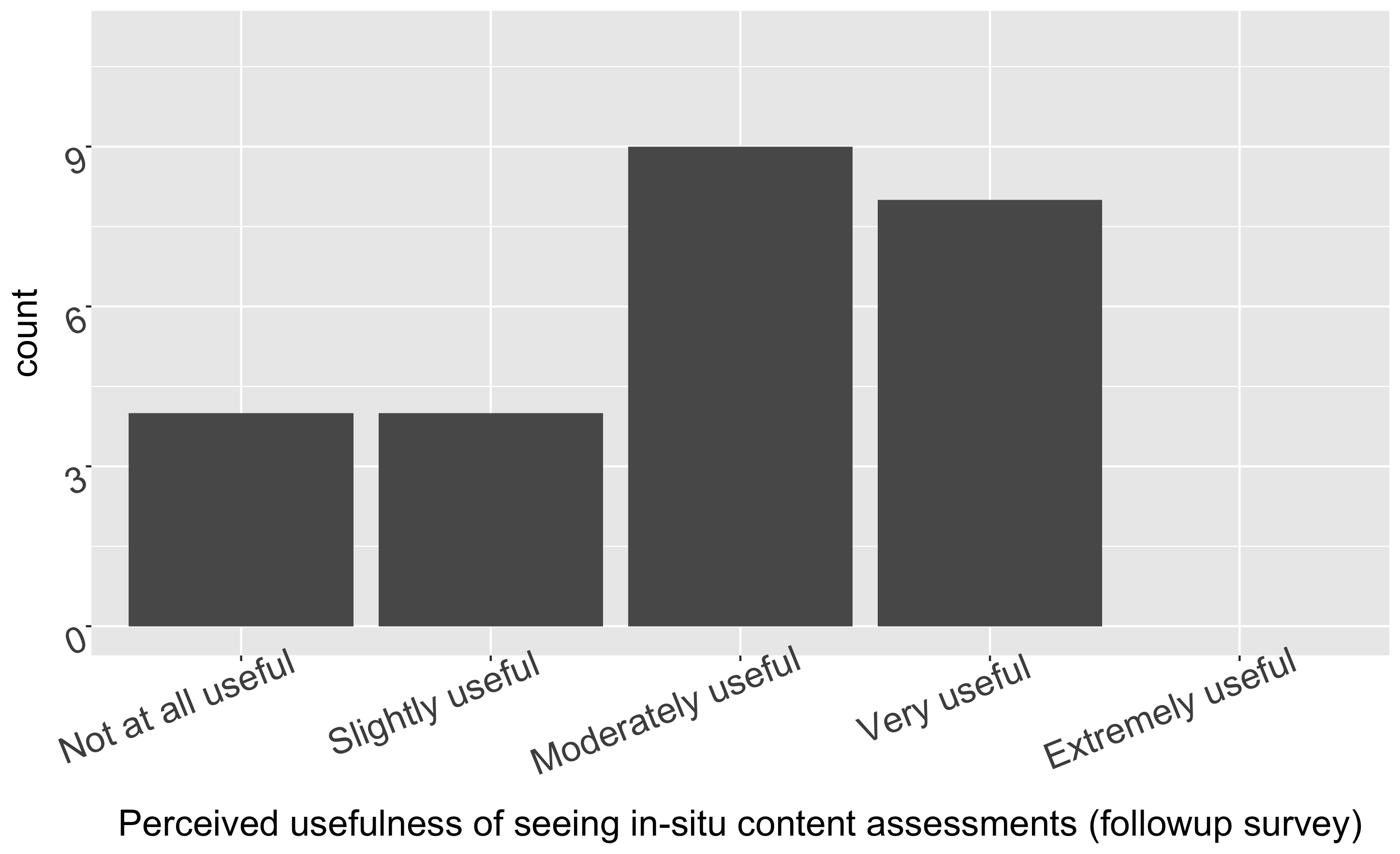}
 \captionof{figure}{The extent to which participants found seeing in-situ assessments helpful on a 1-5 likert scale, in the followup survey.}
 \label{fig:perceived-usefulness-round-2}
  \Description{A bar plot of how much participants found it useful to see in-situ assessments, in the followup survey. All but 4 of the respondents found in-situ assessments at least slightly useful, and 17 out of the 25 respondents found them moderately useful.}
\end{figure}}{
\begin{figure}[!t]
\centering
\begin{minipage}{.47\textwidth}
 \centering
 \includegraphics[width=\linewidth]{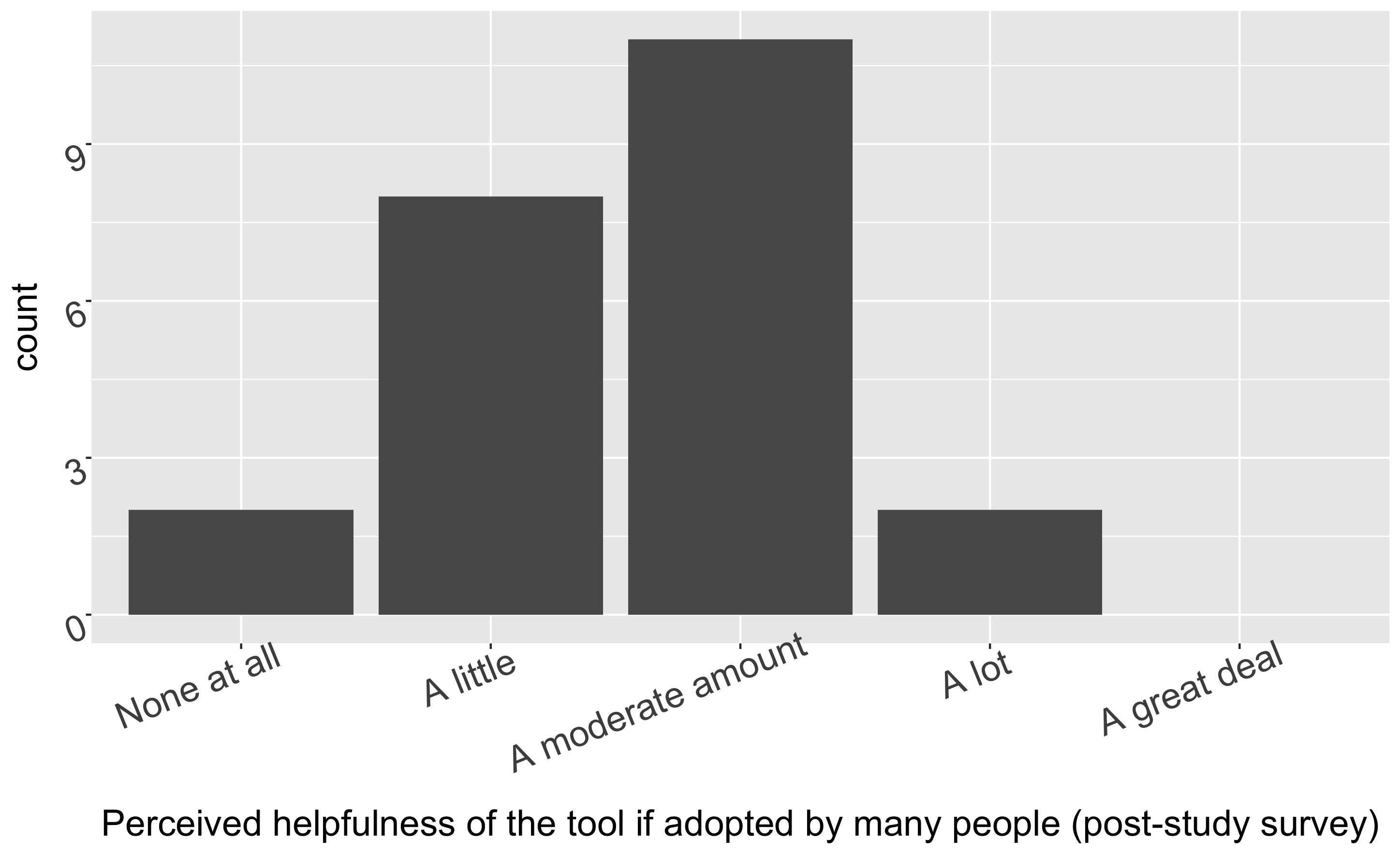}
 \captionof{figure}{The extent to which participants reported the tool would be useful if adopted by many people on a 1-5 likert scale, in the post-study survey.}
 \label{fig:perceived-helpfulness-round-1}
 \Description{A bar plot of how much participants believed the tool would be useful to them if adopted by many people, in the post-study survey. All but 2 of the respondents said the tool would be helpful to them at least a little. 13 out of the 23 respondents found the tool helpful at least a moderate amount.}
\end{minipage}\qquad
\begin{minipage}{.47\textwidth}
 \centering
 \includegraphics[width=\linewidth]{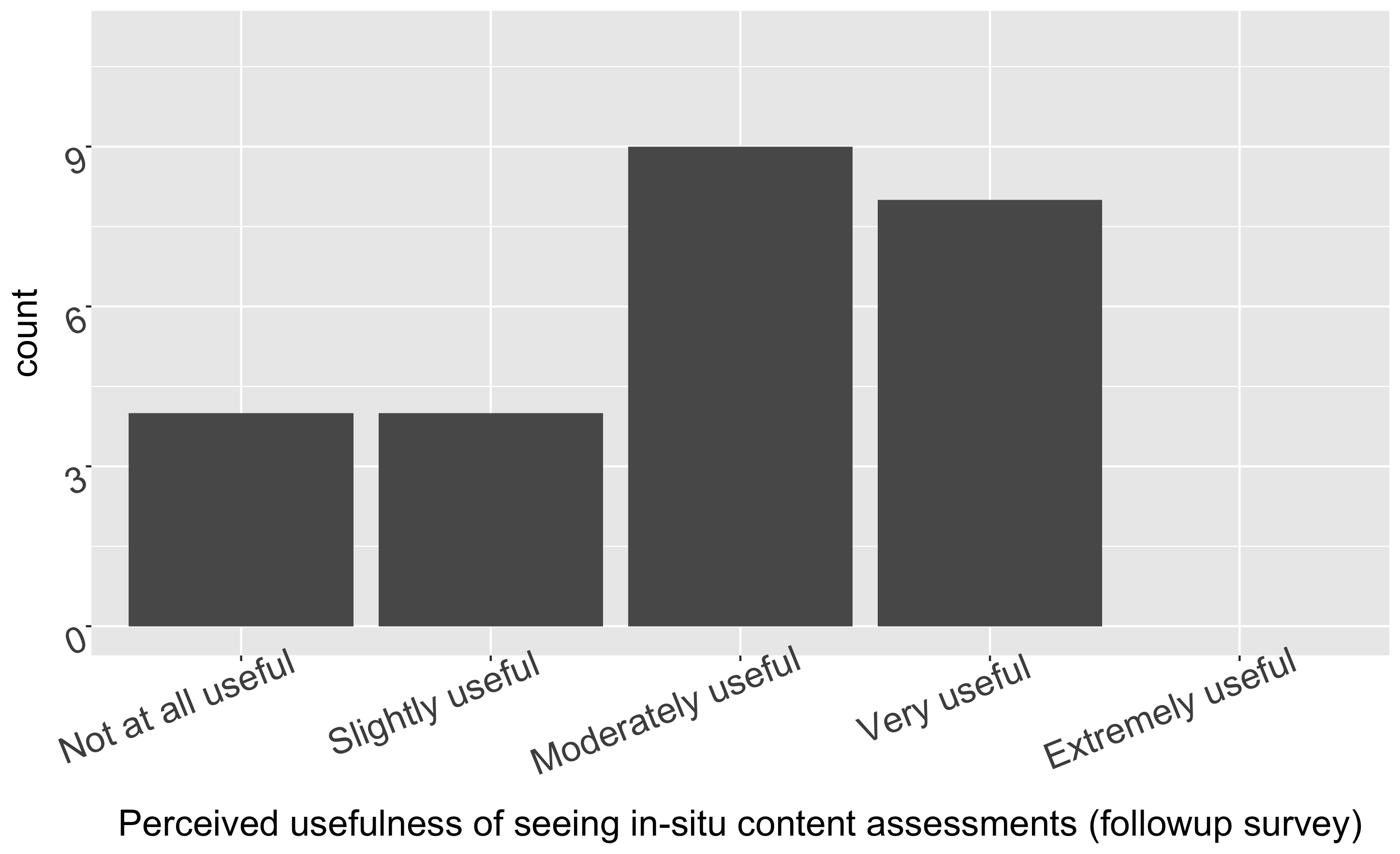}
 \captionof{figure}{The extent to which participants found seeing in-situ assessments helpful on a 1-5 likert scale, in the followup survey.}
 \label{fig:perceived-usefulness-round-2}
  \Description{A bar plot of how much participants found it useful to see in-situ assessments, in the followup survey. All but 4 of the respondents found in-situ assessments at least slightly useful, and 17 out of the 25 respondents found them moderately useful.}
\end{minipage}
\end{figure}
}

\subsubsection{Incentives to Use the Tool, Perceived Problems with, and Ideas for Improving It (RQ3b)}
We asked our participants what would incentivize them to use such a tool, what they would consider as potential reasons they might want to avoid it, and their ideas for resolving those issues. We have summarized their responses in Table~\ref{tab:avoid_incentives}.

A common concern about the tool was that those who post assessments may abuse the ability, e.g., by pushing a certain narrative, using harmful language, engaging in coordinated attacks, etc. Related were concerns about not knowing an assessor's political leanings or agenda. In fact, the Trustnet extension gives users the ability to explicitly specify who they trust and follow and only shows a user assessments from their trusted and followed individuals. The user would not even be aware of assessments on a piece of content by others, even a multitude of bots, if the user does not follow or trust them. The reason why our participants were concerned about this point may have been due to the fact that we set up our study such that all our participants would follow each other by default. Although in our tutorials and our communications with them, we told them that they could unfollow others or choose certain individuals as trustworthy, they may not have paid attention to this point.

In their responses, some participants also mentioned their concerns about specific UI design decisions that we had made. For instance, while some participants found the green flag next to links assessed as accurate helpful, a few said that they would not care as much about if a piece of content is accurate compared to if it is inaccurate. In addition, although some users found the partial fading of content assessed as inaccurate helpful, others were against it or considered it a form of censorship.
These differences in opinion across users point to a need to give users controls for customizing their experience with the actions taken on content once the content has been classified.

\begin{table*}[!t]
\setlength{\aboverulesep}{0pt}
\setlength{\belowrulesep}{0pt}
\setlength{\extrarowheight}{.35ex}
\newcolumntype{g}{>{\columncolor{gray!16}}c}
\setlength{\tabcolsep}{6pt}
\caption{Participants' reasons to avoid the tool and what would incentivize them to use it.}
\Description{Fully described in the text.}
\label{tab:avoid_incentives}
\centering
\begin{tabular}{p{0.3cm}|p{9.5cm}|p{6.5cm}}
  \toprule
  \cellcolor{gray!16}  {} & \cellcolor{gray!16}{Argument} & \cellcolor{gray!16}{Example} \\

  \midrule

     \cellcolor{gray!16} & User abuse including biased assessors, harmful language, assessment bombing of certain authors or  pages, paid experts (N=9) &  \multirow{5}{6.5cm}{
        \textit{``Assessors would have a lot of power. They would need incentives to not also become corrupted (push views of outside parties or their own)...'' (P26)}\vspace{0.1cm}\newline
        \textit{``I'm pretty shy online and I value my privacy and professional life, and wouldn't want anything traced back to me if it could be controversial or seen as too political.'' (P14)}
    }\\
      
    \cline{2-2}
    
    \cellcolor{gray!16} & Time consuming to contribute or read assessments (N=8) & \\
 
    \cline{2-2}
     \cellcolor{gray!16} & Security and privacy concerns (the extension tracking user activity or the user posting assessments publicly) (N=3)  & \\
    \cline{2-2}

\cellcolor{gray!16} & Lack of interest or knowledge in various topics (N=2) & \\
    \cline{2-2}

    \multirow{-8}{*}{\cellcolor{gray!16}\rotatebox[origin=c]{90}{Reasons to avoid}} &  Wanting to develop their own critical thinking, not wanting to be told what to believe (N=2) & 
     \\

  \midrule
  
     \cellcolor{gray!16} & User being knowledgeable and caring about a particular subject, wanting to avoid misinformation (N=10) &  \multirow{8}{6.5cm}{
     \textit{``Some kind of points or a reward system could be an incentive for contributing. If the tool seems useful and valuable, that itself provides a good incentive too...''} (P20)
     \vspace{0.1cm}\newline
     \textit{``Prior to signing up for the platform, if you are going to be someone who assesses, I think it should be displayed on what their political leanings are. If they have certain credentials to assess certain articles or topics, those should be displayed as well. This would give users a better idea on how the assessor is assessing information. It can instill more confidence if they have credentials on certain topics as well. ''} (P17)} \vspace{0.2cm} \\
     
     \cline{2-2}
       \cellcolor{gray!16} & Community engagement, having user's trusted sources use the tool (N=4) & 
    \\
    \cline{2-2}
     \cellcolor{gray!16} & Keeping the assessors in check, e.g., by detecting and removing or blocking bots and abusers, or detecting misinformation in assessments of other users (N=5) & \\

     \cline{2-2}
     \cellcolor{gray!16} &  Monetary compensation (N=4)  & \\
  
     \cline{2-2}
     \cellcolor{gray!16} & Earning reputation points or praise (N=3) & \\
     
    \cline{2-2}
     \cellcolor{gray!16} & The tool should ask and display users' political leanings, biases, and credentials on different topics (N=2) & \\

      \cline{2-2}
     \cellcolor{gray!16} & Having a pool of large and revolving base of experienced assessors (N=1) & \\
     
    \cline{2-2}
       \multirow{-8}{*}{\cellcolor{gray!16} \rotatebox[origin=c]{90}{Incentives}} & Restricting the showing of assessments to topics of interest (N=1) &
     \\
  
  \bottomrule
\end{tabular}
\end{table*}

\section{Discussion and Future Work}
The universal in-place assessments displayed through the Trustnet browser extension would be in contrast to expecting the user to actively seek fact-checking information from external sources for every piece of content that they encounter. Furthermore, the extension enabling the user to provide assessment in-place can help others who find the user a trustworthy source.

We observed that our users used the extension to assess a variety of different types of content on various websites, blogs, and social media, and on various topics not limited to politics or health. Not all sources were widely known, but they still have enough consumers for the accuracy of the content they publish to be consequential.
In the existing fact-checking model on today's web, the power to assess content is confined to a limited set of individuals and fact-checking initiatives who do not have the resources to cover every piece of content published by every source. For instance, a number of Youtube channels whose videos our participants had assessed as well as some specialty websites and blogs had enough subscribers for a wide reach but perhaps not enough to garner attention from the fact-checking organizations. To scale fact-checking in a world where everyone could be a publisher, we need all the help we can get, even from regular users and perhaps especially those who have interest in and knowledge about a particular topic, to sieve accurate from inaccurate information.

While assessments in our tool are democratized, how often a user views content with credibility assessments depends on who and how many sources they have marked as trustworthy. The issue of scaling such democratized trusted assessments so that a user sees assessments on content more often can be dealt with in two ways. One is using an AI that learns assessments from a select set of people (e.g., a user's trusted associates) and predicts how they would assess other similar content, such as the AI in~\cite{jahanbakhsh2023AI}. Another is to explore whether a more extensive trust network can be built for each user by leveraging transitivity of trust. Prior work has also examined trust transitivity in social networks, in the context of e-commerce, recommender systems, and chat moderation~\cite{liu2011trust, josang2003analysing, cobleigh2020trustnet}. Indeed, there is also the potential for a hybrid approach where the extension delivers professional fact-checking information to users in addition to that from regular users, but leaves it to the users to decide which fact checkers to trust.

\subsection{Design Implications}
\subsubsection{User Criteria for Evaluating Content Accuracy}
We reported on the criteria that our users used for assessing content credibility in Section~\ref{section:criteria}. In contrast to other studies that have presented criteria for assessing content credibility~\cite{zhang2018structured} and \cite{jahanbakhsh2021exploring}, in our study, users were not restricted to a set of curated news claims or articles, but could use the tool to submit their assessment of any piece of content of their choosing. In this in-the-wild setting, some of the rationales that emerged out of the assessments encompassed, but were broader than, those previously reported in~\cite{jahanbakhsh2021exploring} and~\cite{zhang2018structured}, and some had not been reported before.

Tools that allow users to evaluate content credibility, such as ours, can be augmented with a checklist of criteria surfaced in our study as a form of guideline to nudge the user to reflect on various aspects of credibility, when sharing content, similar to the proposal in~\cite{jahanbakhsh2021exploring}, or even when reading content. While in~\cite{jahanbakhsh2021exploring}, going through the checklist in addition to providing free-text reasoning at the time of sharing did not lead users to share less misinformation compared to providing free-text assessments alone, future work should study whether a different set of criteria such as ours could be effective. Even if the structured checklist does not reduce the sharing of misinformation, it can still help those who view the criteria that the assessors have highlighted in structured form, and perhaps even use them in filtering through content or assessments.

\subsubsection{Incentivizing Participation}
A hurdle in using the tool that participants mentioned is that it can be time consuming to contribute or read assessments. Future work can investigate how to further aid users in their sensemaking process, e.g., by providing a mechanism using which they can collectively cluster, synthesize, and surface unique information in the assessments, similar to Crowdlines or Wikum~\cite{zhang2017wikum, luther2015crowdlines}. The tool can then visualize the analysis provenance so that other users can audit the sensemaking pipeline~\cite{li2020crowdtrace}. The assessments pane can also be structured to include a section into which users can clip ``evidence'', including links to public hearing videos, studies, fact-checking articles, etc. that can help establish the veracity of claims in the content.

Our users also desired some form of compensation or recognition for their effort. The tool can be enhanced with a reputation system similar to the one in Stack Overflow where users accrue points based on how others rate their assessments or questions about content accuracy. To protect against coordinated attacks against an individual, the reputation score a user A views for user B would not be a global score, but rather a function of the scores that A's trusted sources (or those who have an extended trust path to A) have given to B's assessments. In fact, the score system can be used to incentivize users to write assessments that can reach across the political divide or to read assessments from others that are not similar to them, e.g., by assigning a higher weight to the upvote on an assessment from an out-group user compared to those from the in-group users. 

\subsubsection{Democratizing the Design Space of Actions on Labeled Content}
We scoped our approach and consequently our tool to democratizing the power to categorize the accuracy status of content. Informed by literature, we then made specific design decisions about what to do with the content that was labeled as accurate or inaccurate and how to signal its status. Our users however, were not in agreement about whether they liked these specific design decisions. For example, while the accuracy and inaccuracy of content were of equal importance to some, others wished to only be notified of inaccuracies. These differences make the case for democratizing the choice in the design space of actions as well.
In fact, this is another aspect of moderation that can be democratized regardless of whether the arbitration itself is. Even if accuracy labels are generated by a select few, the decision of whether, how, and in what circumstances users want to view labels can be deferred to users.

\subsection{Content Moderation on the Client Side}
In the current information ecosystem and without much control over choosing what content they wish to see or avoid, users have submitted to the will of the platforms and their feed curation algorithms. The power of what information flows where is much tilted in the favor of the platforms. Our work is a case study of how we can give more agency to the users and democratize content moderation on the web, without support from the underlying platforms. While our tool was focused on enabling users to assess the accuracy of content and see assessments from their trusted sources, the system can be generalized, empowering users in a variety of other use cases. For instance, our users desired to see assessments from users or news outlets with diverse perspectives or from different sides. Taking this idea further, the architecture and design of our tool can be used to enable users to provide alternative sources for content--- those with a different political leaning similar to an in-situ AllSides~\cite{allsides}, those that a user thinks are more credible than the source at hand, or those that together with the source can paint a more complete picture of the story.
The system can be adopted to label content that is hostile to various marginalized identities, similar to the Shinigami Eyes browser extension that colors trans-friendly and anti-trans social media groups, users, and search results~\cite{shinigami}. The system can be used to label content with certain valence, for instance ``depressing'' or ``angry'', so that the user may choose to filter out labeled content if and when they are not in a state of mind to consume it. Future work can study different types of democratized governance on these tools, e.g., leaving the decision of categorization to customized trusted sources specified by the user as in our study or allowing the user to choose among various sets of moderator bodies with different standards for what constitutes a positive label. The choice of whether such content should removed from the user's view can be given to the user as well. Indeed, in our study, not every user was in agreement about whether visual fading of content assessed as inaccurate is an acceptable design decision. 

The user-generated data in such a tool can further be augmented with machine learning or user-configurable word filters similar to~\cite{jhaver2022designing} to block content on certain topics, with certain valence, or from certain people that appears on the user's feed. While users cannot stop the flow of unwanted and unasked for messages, for instance those that are forcefully boosted to them (e.g., Elon Musk's tweets receiving the ``power user multiplier'' special treatment~\cite{elonMuskSpecialTweets}), they can leverage browser extensions such as ours to locally flag such content or remove it from the rendered page.

\subsection{Broader Social Implications}
\subsubsection{Are regular users able to find trustworthy sources?}
The ecosystem that we propose relies on the users to find and vet others who should be trusted to provide assessments. But is this a challenge that could render the approach not feasible? We argue that on the contrary, this approach enables users to do online what they would do in natural settings, as finding trusted sources and heeding the information that they provide is how people have always determined the truth. Epistemically, social construction of knowledge is individuals receiving testimony from trusted sources about information that they cannot verify by direct observation~\cite{steup2005stanford}. On social platforms, users commenting about a post’s accuracy contribute to the socially constructed knowledge of others who trust them~\cite{bruckman2022should}. Additionally, the presupposition of democracy is that we recognize our fellow citizens as having the capacity to participate in collective governance.
However, social platforms, or the web, currently do not accommodate these age-old practices well.

\subsubsection{Echo Chambers}
An adverse effect of giving users autonomy to decide from whom they would like to see assessments can be the potential for echo chambers. But does this mean that instead, a governing body should make decisions that it decrees are in the best interest of users?
Prior work has warned against relinquishing the power of content moderation to platforms and advocated for user autonomy in this space and argued that individuals have a moral right to freedom of listening and against compelled listening~\cite{jahanbakhsh2022leveraging, jahanbakhsh2023AI}. These rights are delivered to users on some platforms through affordances that allow them to for instance, tag, rate, or upvote content and filter based on this metadata. Therefore, such affordances already have precedence on the web. Another advantage to structuring trust and assessments on the web is that such tools as ours can deliver features aimed at directing users to trustworthy contrary viewpoints. In fact, prior work has also reported that users are more receptive to fact-checking information from friends than strangers~\cite{hannak2014get, margolin2018political}.
In our tool, we made an initial attempt at encouraging users to expand their network by having the extension inform the user if there exist assessments on the page by someone that they do not follow or trust. Because of the sparsity of the data in our small-scale user study, the recommended sources on a page were those who were trusted by the most number of users platform-wide. However, with a larger adoption of the tool, the extension can recommend to a user contrary viewpoints by sources that for instance, have a history of assessments that align with the user's, or sources that are trusted by those that user deems trustworthy.

\subsection{Challenges of Finding Link Redirects and Possible Solutions}
To display user-generated content, for many use cases including ours, it is necessary to find embedded links on the page that the user is visiting as well as to which resource each link redirects. 
A potential complication of the client following the trail of redirects for many links from the same domain is that the client can get rate-limited. In our system architecture, because the server caches the redirects that the clients have found, with a large enough number of users, the number of links for which each client needs to find the redirected resource becomes fewer. 
To further protect against getting rate-limited, the client does not fetch all the links it finds on the page at once, but rather batches the requests and spreads them out. We set the delay between the request batches the same across all websites because many websites do not disclose their rate limits. This static delay results in suboptimal delay in fetching the links on some domains that would allow for more frequent requests and conversely, getting rate-limited on others that only allow for less frequent requests.

However, this issue can be dealt with in two ways. The first is to keep an estimate of the rate-limit of each domain on the server and dynamically change it until the estimate appropriately captures the domain's actual rate-limit. This can be done by for instance, additively increasing the rate of requests the client makes on the domain if it does not get rate-limited, and conversely, multiplicatively decreasing the rate of requests if it does, similar to congestion control schemes in networks. The second is to learn whether the links to internal and external resources on a website are likely to go through redirections. While the extension currently attempts to follow all links found on a webpage, the links on many websites in fact---at least those that direct to internal resources---do not go through redirections. Identifying these websites after a number of encounters can prevent the extension from getting rate-limited, decrease the delay of fetching user-generated content on the links, and reduce the load of caching mappings of links on the server. 


\section{Limitations}
A limitation of the empowerment brought about by our tool is the inherent limitation of browser extensions---that they only work on desktop and not on mobile browsers or within mobile applications. This is unfortunate especially because certain messaging apps such as WhatsApp have been a hotbed of misinformation~\cite{varanasi2022accost}. To partially address this issue, we can provide databases or platforms such as~\cite{jahanbakhsh2022leveraging} where users can copy and paste in-app messages or links to assess them or see their assessments. However, this solution relies on users adopting this extra step into their content consumption process.

Another limitation of this tool concerns how Facebook deals with links. When displaying posts or external links embedded in posts (e.g., posts from news media), the rendered DOM does not include their URLs, unless as a result of user action such as the user hovering their mouse over the elements. Therefore, without this extra action from the user, the extension cannot know the URLs of the displayed posts or content to fetch their assessments. Interestingly, embedded links used to have their URL in the DOM in the early days of the Trustnet extension, but no longer do. The rationale behind this change in design is unclear. However, it points to the need for active maintenance of browser extensions such as Trustnet to keep working on hostile platforms.

While it is conceivable to view a source as generally well-informed in their assessments, trust in a source may not always be a simple binary choice of yes or no. Future work should study how users reason about their degree of trust in various sources and how to incorporate that into such a tool as ours.

\section{Conclusion}

In this work, we explore democratizing misinformation moderation on the web by providing users with a browser extension that empowers them to assess any content including news articles, social media posts, or videos, and to see assessments from the sources they have marked as trustworthy in-situ. We evaluated the potentials and the challenges of such a design through a two-week user study where we asked users to use the tool to assess any content that they wanted. We observed a gap between the content that users want to assess and whose accuracy they consider important and the type of content that is usually assessed by fact-checking initiatives, pointing to a need for giving users autonomy in this space. Our users perceived value in the ability to assess content or see assessments from others right on the content, for instance, because they believed the process helped them think critically about news. They also had ideas for improving the process such as incorporating a reputation or a point system to incentive users to contribute assessments. We also investigated user's rationales for their assessments and believe they can extend credibility indicators reported in prior work and can be used for guiding users in how to evaluate content accuracy. Our work suggests there is potential for democratizing content moderation on the web through tool such as ours.

\begin{acks}
We thank all the alpha testers of our tool as well as the users of our study. In addition, we thank Professor Michael Bernstein for his feedback on the manuscript.
\end{acks}

\bibliographystyle{ACM-Reference-Format}
\bibliography{bibliography}

\appendix

\section{Participation}
\label{appendix:participation-plot}

\begin{figure}[t]
  \centering
 \centering
   \includegraphics[width=\linewidth]{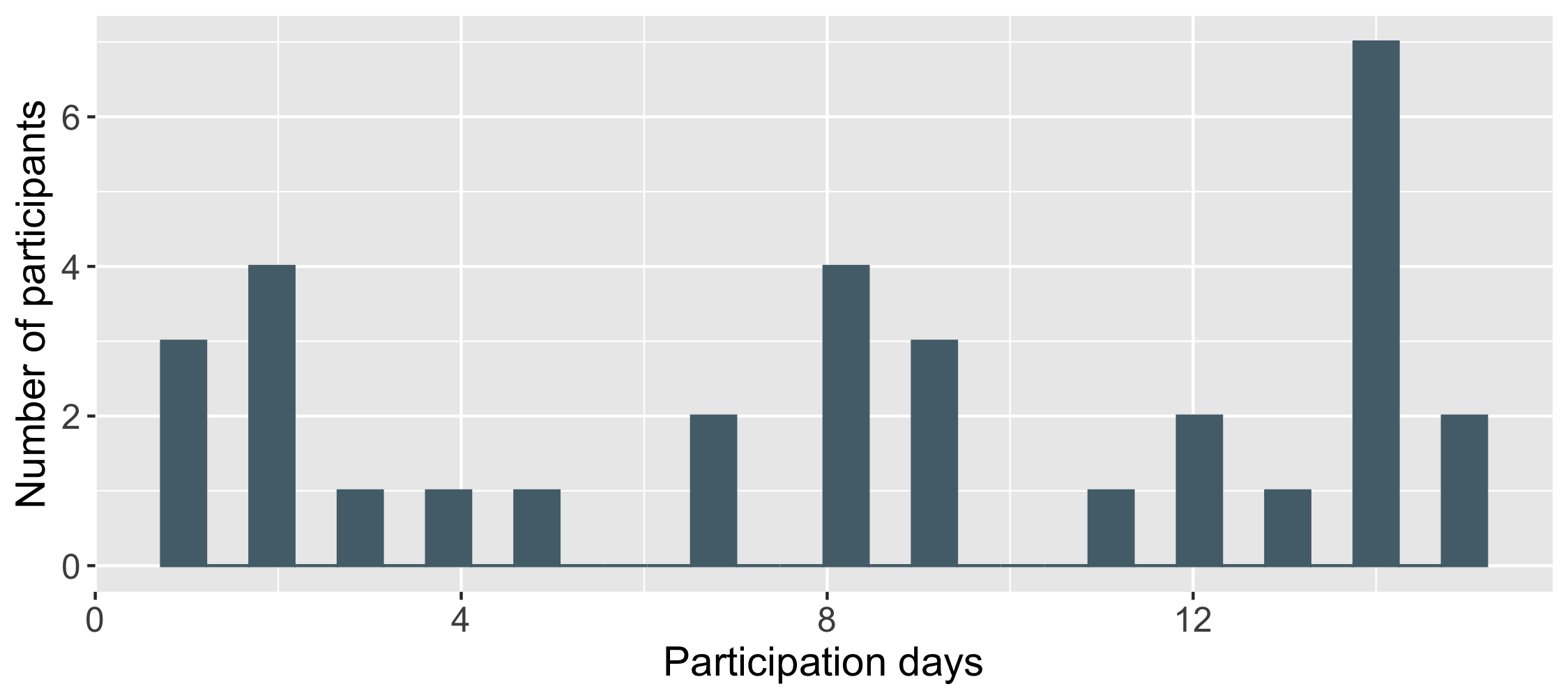}
  \captionof{figure}{The distribution of how many participants completed the daily tasks for how many days. Not all the ones that completed the tasks for fewer than 14 days dropped out before the study's conclusion. Rather, they did not complete the tasks in certain days.}
 \label{fig:particpation-plot}
 \Description{The plot shows that more participants completed the daily tasks for more than a week than less. The median of how many days our participants completed the daily tasks is 8.5.}
\end{figure}

Figure~\ref{fig:particpation-plot} shows the participation distribution across our study participants.

\section{Partisanship}
Prior work has reported that users' assessments of content accuracy correlates with whether the their partisanship aligns with that of the content~\cite{jahanbakhsh2021exploring, allen2022birds}. We wanted to further investigate whether partisanship plays a role in the context of our study where users were not restricted in their choice of which content to assess.
Therefore, we consulted the media bias ratings provided by 3 sources (AllSides~\cite{allsides}, Ad Fontes Media~\cite{adfontesmedia}, and Media Bias/Fact Check (MBFC)~\cite{mediabiasFactcheck}) and developed a normalized numeric political leaning value averaged from the ratings given by these sources. The value ranged from -1 to 1, with  negative values indicating a left-leaning ideology, and positive values indicating a right-leaning ideology. A number of domains in our dataset were aggregator websites such as Hacker News, Allsides, and Yahoo News that present articles from various media. For retrieving the bias of the articles assessed by our users on these aggregators, we considered the original source of the articles, for instance \textsf{LA Times} for an article assessed on Yahoo News.

The political leaning information on a number of sources was not available in these repositories. Some of these sources for instance, were not major news websites or covered specialty news. In fact, the lack of information about bias or credibility of a number of sources that our small user sample consumed points to the gap---perhaps resulting from limited resources---between the content fact-checked by professionals and the content that the general public consumes and about whose accuracy they inquire. The number of datapoints (i.e., user assessments) for which we had the political leaning of the media source was 588.

To measure partisanship concordance between a user and the content that the user assessed, we made the simplifying assumption that the leaning of the content is the same as the leaning of its media source. We labeled each user as either a Democratic or a Republican. We were able to place all participants including those who identified as Independent or other in one of the two Democrat or Republican categories because in addition to party, we had asked participants about their political preference (strongly Republican, lean Republican, Republican, Democrat, lean Democrat, strongly Democrat). If the numeric political leaning value of the media source was negative and the user was labeled as a Democratic, or if the media value was positive and the user was a Republican, partisanship concordance had a value of 1. Otherwise, partisanship concordance was 0. 

To understand if partisanship concordance has an effect on users' accuracy judgment, we fit a generalized linear model to the data. To do so, we used the function "glmer" from the R package "lme4" and used the family function "Binomial" with the link "logit". We included user and content identifiers as random effects to account for variation in the outcome as a result of the unobserved characteristics of a particular user or a piece of content. The model was as follows:


\begin{align}
    \label{equation:assessment_concordance}
    \text{accuracy assessment}_{uc} &= \beta1(\text{ partisanship concordance})_{uc} \nonumber \\
    &\quad+ b_{u}Z_{u} + b_{c}Z_{c} + \epsilon_{uc}
\end{align}

where the $\text{accuracy assessment}_{uc}$ is user $u$'s assessment of content $c$. The indicator variable $1(\text{partisanship concordance})_{uc}$ denotes whether user $u$'s political leaning is aligned with content $c$. 
$Z_{u}$ is the matrix for the random effects for observations for participant $u$, and $Z_{c}$ for content $c$. $\epsilon_{uc}$ is the error term. $\beta$ is the coefficient of interest.

Interestingly, we observed that partisanship concordance did not have a significant effect on users' assessment [$\beta=0.45, z=0.56, p=0.58$], suggesting that the effect of partisanship on users' assessments when users are not restricted in what content to assess warrants more research.

\end{document}